\documentclass[11pt,twocolumn]{article}

\usepackage{amsmath,amsfonts,amssymb}

\usepackage{float}

\usepackage[nolist]{acronym}

\usepackage{psfrag}		
\usepackage{epstopdf}	
\usepackage{pgfplots}

\usepackage{graphicx}

\usepackage{tikz,ifthen,tikz-3dplot}

\usepackage{subfigure}

\usepackage{enumitem}
\usepackage{float}

\usepackage{dsfont}

\DeclareGraphicsExtensions{.pdf,.jpeg,.png,.eps}
\graphicspath{{./graphics/}}

\definecolor{comment}{RGB}{190,10,20}

\makeatletter										%
\def\blfootnote{\xdef\@thefnmark{}\@footnotetext}	
\makeatother										%
\renewcommand{\Vec}[1]{\boldsymbol{#1}}
\newcommand{\Mat}[1]{\boldsymbol{#1}}

\newcommand{\delete}[1]{}

\newcommand{\trace}[1]{{\mathrm{t}}\mathrm{race}\ #1}
\newcommand{\Diag}[1]{{\mathrm{D}}\mathrm{iag}\ #1}
\newcommand{\diag}[1]{\mathrm{d}\mathrm{iag}\ #1}
\DeclareMathOperator*{\argmin}{arg\,min}
\DeclareMathOperator*{\argmax}{arg\,max}

\newcommand{\be}{\begin{equation}}
\newcommand{\ee}{\end{equation}}
\newcommand{\tempOff}[1]{}

\begin{document}

\title{ Exploring the Physical Layer Frontiers of Cellular Uplink - The Vienna LTE-A Simulator \thanks{submitted to Eurasip Journal on Wireless Communications and Networking on 07-Sep-2015, Manuscript ID: JWCN-D-15-00369} }

\maketitle

\author{Erich Z\"ochmann, \and Stefan Schwarz, \and Stefan Pratschner, \and Lukas Nagel, \and Martin Lerch, \and and Markus Rupp \\\thanks{Institute of Telecommunications, TU Wien}}


\blfootnote{This work has been funded by the Christian Doppler Laboratory for Wireless
Technologies for Sustainable Mobility, KATHREIN Werke KG, A1 Telekom
Austria AG and the Institute of Telecommunications, TU Wien.  The financial support by the Austrian Federal Ministry of Economy, Family and Youth and the National Foundation for Research,
Technology and Development is gratefully acknowledged.}

\begin{abstract}
Communication systems in practice are subject to many technical/technological constraints and restrictions. MIMO processing in current wireless communications, as an example, mostly employs codebook based pre-coding to save computational complexity at the transmitters and receivers. In such cases, closed form expressions for capacity or bit-error probability are often unattainable; effects of realistic signal processing algorithms on the performance of practical communication systems rather have to be studied in simulation environments. The Vienna {LTE-A} Uplink Simulator is a 3GPP {LTE-A} standard compliant link level simulator that is publicly available under an academic use license, facilitating reproducible evaluations of signal processing algorithms and transceiver designs in wireless communications. This paper reviews research results that have been obtained by means of the Vienna LTE-A Uplink Simulator, highlights the effects of Single Carrier Frequency Division Multiplexing (as the distinguishing feature to LTE-A downlink), extends known link adaptation concepts to uplink transmission, shows the implications of the uplink pilot pattern for gathering Channel State Information at the receiver and completes with possible future research directions. 
\end{abstract}

\section{Introduction}

Current cellular wireless communications employs Universal Mobile Telecommunications System (UMTS) Long Term Evolution (LTE) as the high data rate standard~\cite{3GPP_TS_36211}. The increasing demand of high data traffic in up- and downlink forces engineers to push the limits of {LTE}~\cite{Sch13}, e.g., through enhanced multi-user Multiple Input Multiple Output (MIMO) support~\cite{Yuyu2012,Pan2014\tempOff{,Schwarz-TWC2014}}, Coordinated Multipoint (CoMP) transmission/reception~\cite{Falconetti2011,Qiang2013\tempOff{,Sch14b}} as well as improved Channel State Information (CSI) feedback algorithms~\cite{\tempOff{Sch13b,}Xue2014}. The authors of \cite{5Gdahlman} predict further evolution of existing {LTE}/ LTE-Advanced (LTE-A) systems in parallel to development of new radio-access technologies operating at millimetre wave frequencies until 2020 and beyond. Fair comparison of novel signal processing algorithms and transceiver designs has to assure equal testing and evaluation conditions to enable reproducibility of results by independent groups of researchers and engineers~\cite{ViennaLTESimulators}. For performing system-level simulations \cite{baldo2011open}, \cite{piro2011simulating} or \cite{stea2014simulte} are freely accessible options. For link level there are mainly commercial products available that facilitate reproducible research, such as, {\it is-wireless} LTE~PHY~LAB \cite{iswirelessphylab} or {\it Mathworks} LTE System Toolbox \cite{lte_matlab}. To the best of the authors knowledge, however, the Vienna {LTE} simulators are the only suite of simulation tools for {LTE} system and link level publicly available under an academic use licence, thus, free of charge for academic researchers all over the world. In this paper, we introduce the latest member of the family of Vienna {LTE} simulators, that is, the Vienna {LTE-A} uplink link level simulator, and highlight our research conducted by means of this simulator.

The outline of this article is as follows: We start with a brief re-capitulation of the {LTE-A} specifics and introduce the modulation and multiple access scheme and the employed MIMO signal processing  of {LTE-A} uplink in Section \ref{sec:system}. We then develop a matrix model describing the input-output relationship of the {LTE-A} uplink and present Signal to Interference and Noise Ratio (SINR) expressions for Single Carrier Frequency Division Multiplexing (SC-FDM) as well as Orthogonal Frequency Division Multiplexing (OFDM). The {OFDM} {SINR} expression and the performance of OFDM will serve as reference to study to effects of DFT-spreading imposed by {SC-FDM}. 

In Section \ref{sec:effects}, we investigate the physical layer performance of {SC-FDM} and {OFDM}, comparing Bit Error Ratio (BER) and Peak to Average Power Ratio (PAPR). {BER} for LTE SISO transmissions were already analysed in link-level simulations by  \cite{zhang2006comparison,sinanovic2011comparison,suarez2012lte} and semi-analytically by \cite{BER_SCFDMA,sanchez2011ber}. By means of our simulator, we reproduce these results and provide bounds to predict the performance of {SC-FDM} with respect to {OFDM}. The insights gathered by the BER simulations allow us to interpret the difference in throughput obtained by OFDM and SC-FDM, as discussed in Section \ref{sec:adaptation}. 

Based on the {SINR} expressions developed in Section \ref{sec:system}, we present a limited feedback strategy for link adaptation in Section \ref{sec:adaptation} and contrast the performance of {LTE} uplink with channel capacity and other performance upper bounds that account for practical design restrictions~\cite{Schwarz-bounds2011}. Until Section \ref{sec:CE_reference_symbols} we assume perfect {CSI} at the receiver. The remaining sections will describe methods to obtain {CSI} at the receiver.

In Section \ref{sec:CE_reference_symbols}, we highlight and describe the Demodulation Reference Signal (DMRS) structure employed in {LTE-A} uplink to facilitate channel estimation of the time-frequency selective wireless channel.
 
Based on the obtained insights, we elaborate on the basic concept of Discrete Fourier Transform (DFT)-based time domain channel estimation in Section \ref{sec:CE_channel_estimation} and review alternative code / frequency domain methods that can outperform {DFT}-based schemes~\cite{pratschner2015uplinkCE}. 

Due to the increasing number of mobile users that stay connected while travelling in cars or (high speed) trains, we then shift our focus to high velocity scenarios. Such scenarios entail high temporal selectivity of the wireless channel, rendering accurate channel interpolation very important to sustain reasonable quality of service. We introduce and investigate basic concepts of channel interpolation in Section \ref{sec:channel_interpolation}. 

We briefly discuss open questions for future research in Section \ref{sec:Future} and conclude in Section \ref{sec:Conclusion}. Details to the handling of the simulator are provided in \cite{sim_doc}.

\subsection*{Notation}

Matrices are denoted by bold upper-case letters such as $\Mat{H}$ and vectors by bold lower-case letters such as $\Vec{h}$. The entries of vectors and matrices are accessed by brackets and subscripts, e.g., $[\Vec{h}]_k$ and $[\Mat{H}]_{k,n}$. Spatial layers or receive antennas are denoted by superscripts in braces, e.g., $\Vec{x}^{(l)}$. The superscripts $(\cdot)^T$ and $(\cdot)^H$ express transposition and conjugate transposition. ${\| \cdot \|_2}$, $\| \cdot \|_\infty$ and $\| \cdot \|_F$ symbolizes the Euclidean-, the Maximum- and the Frobeniusnorm, respectively. The entrywise (Hadamard) product is denoted by $\odot$ and the Kronecker product by $\otimes$. The all ones vector/matrix is denoted by $\mathds{1}$. The operator $\Mat{X}=\operatorname{Diag}(\Vec{x})$  places the vector $\Vec{x}$ on the main diagonal of $\Mat{X}$ and conversely the operator $\Vec{x}=\operatorname{diag}(\Mat{X})$ returns the vector $\Vec{x}$ from the main diagonal of $\Mat{X}$. A block-wise Toeplitz (circulant, diagonal) matrix is a block matrix with each matrix of Toeplitz (circulant, diagonal) shape. The size of matrices is expressed via their subscripts, whenever necessary. \delete{\\ 
Frequently used matrix symbols are: $\Mat{H}$ for channel matrices, $\Mat{I}$ for the identity matrix, $\Mat{D}$ as {DFT} matrix, $\Mat{S}$ as block selection matrix and $\Mat{R}$ for the {DMRS}. There are three basis sets for the layer index $l \in \lbrace 1, \dots, L \rbrace$, the frequency index $k \in \lbrace 1,\dots, K  \rbrace$ and the time index $n \in \lbrace 1, \dots, N \rbrace$.}

\begin{figure}
	\begin{center}
		\begin{tikzpicture}[>=stealth,scale=0.45, every node/.style={scale=0.7},
	 light/.style={thin},
	 grid/.style={very thin,gray},
	 myaxis/.style={line width=1mm}]

\usetikzlibrary{decorations.pathreplacing} 

\draw[white] (-3,-4) rectangle (10,8);

\draw[light] (0,-0.35) 	-- 	(0,-3.4);		
\draw[light] (3.5,-0.1) -- 	(3.5,-2.3);		
\draw[light] (7,-0.1) 	-- 	(7,-3.4);		

\draw[<->,light] (0,-2.1) -- (3.5,-2.1) 	node[midway,above]{slot};
\draw[<->,light] (0,-3.2) -- (7,-3.2)		node[midway,above]{subframe};

\draw[grid,step=0.5cm] 	(0,0) grid 			(7,6);
\draw[very thick] 		(4,4.5) rectangle 	(4.5,5);
\draw[very thick] 		(0,0) rectangle 	(3.5,6);
\draw[->, thick] 		(-0.25,0) -- (8.5,0);
\draw[->, thick] 		(0,-0.25) -- (0,7.5);

\draw					(3.5,-0.6) 				node[anchor=north,fill=white]{symbols $n$};
\draw					(0,3) 					node[anchor=south,rotate=90]	{subcarrier $k$};

\draw[thick,white] (7.3,0) -- (7.8,0);
\draw[thick,fill=white, line cap=round, dash pattern=on 0pt off 2\pgflinewidth] (7.3,0) -- (7.8,0);

\draw[thick,white] (0,6.3) -- (0,6.8);
\draw[thick,fill=white, line cap=round, dash pattern=on 0pt off 2\pgflinewidth] (0,6.3) -- (0,6.8);

\draw (0.25,0) node[anchor=north]{0};
\draw (3.25,0) node[anchor=north]{6};
\draw (3.75,0) node[anchor=north]{0};
\draw (6.75,0) node[anchor=north]{6};
\draw (0,0.25) node[anchor=east]{0};
\draw (0,5.75) node[anchor=east]{11};

\draw[-latex] (4.5,7.5)	node[anchor=west] {Resource Block} 		to [out=180,in=90] (2.75,6);
\draw[-latex] (5.5,6.5) node[anchor=west] {Resource Element} 	to [out=180,in=90] (4.25,5);

\end{tikzpicture}
		\caption{The LTE-A uplink resource grid.}
		\label{fig:resource_grid}
	\end{center}
\end{figure}
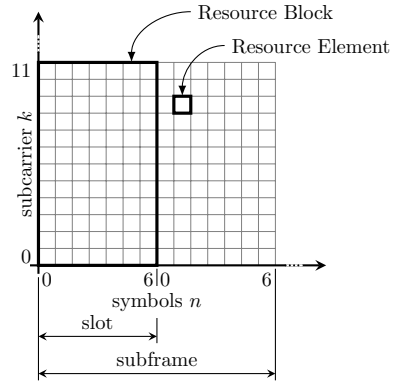

\section{LTE-specific System Model and SINR}
\label{sec:system}

\begin{figure*}[ht!]
	\includegraphics[width=0.98\textwidth]{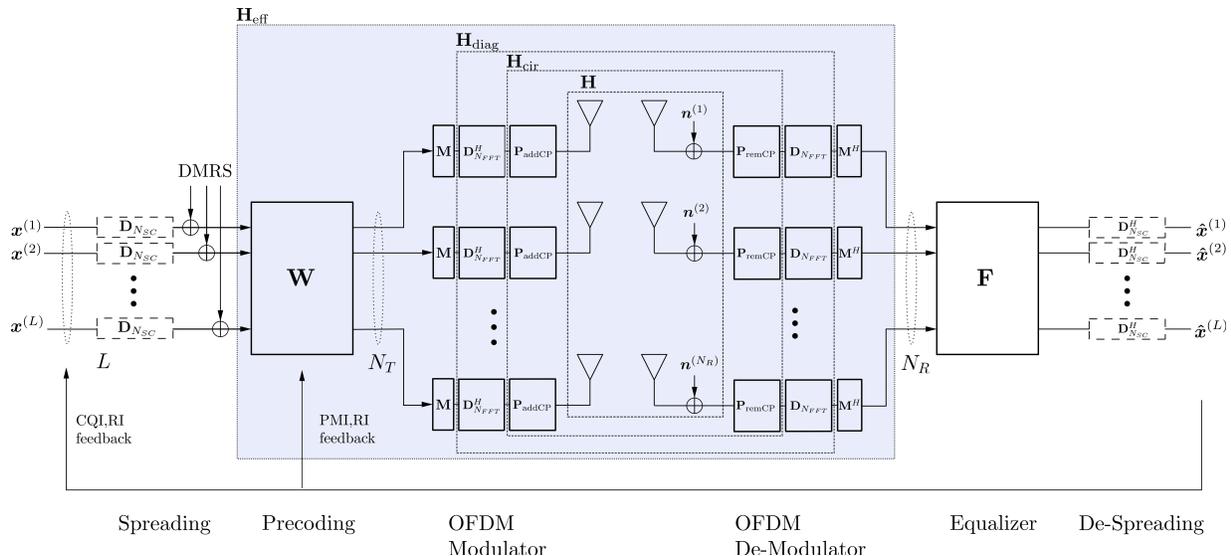}
\caption{The LTE-A uplink transceiver.}
	\label{fig:system_model}	
\end{figure*}
\begin{table*}

 \begin{align}
 \Vec{\hat{x}} =& \left( \Mat{I}_{L} \otimes \Mat{D}_{N_{SC}}^H \right)  \Mat{F} \left( \Mat{I}_{N_{R}} \otimes  \Mat{M}^H \Mat{D}_{N_{FFT}} \Mat{P}_{\text{remCP}}\right)  \Mat{H}  \left( \Mat{I}_{N_{T}} \otimes  \Mat{P}_{\text{addCP}}\Mat{D}_{N_{FFT}}^H \Mat{M}\right)\left( \Mat{W} \otimes \Mat{I}_{N_{SC}} \right)\left( \Mat{I}_{L} \otimes \Mat{D}_{N_{SC}} \right)\Vec{x} \nonumber \\ & \hspace{236pt}+ \underbrace{\left( \Mat{I}_{L} \otimes \Mat{D}_{N_{SC}}^H \right)  \Mat{F}\left( \Mat{I}_{N_{R}} \otimes  \Mat{M}^H \Mat{D}_{N_{FFT}} \Mat{P}_{\text{remCP}}\right) \Vec{n}}_{\tilde{\Vec{n}}} \nonumber \\[-30pt]  
= & \overbrace{\big( \Mat{I}_{L} \otimes \Mat{D}_{N_{SC}}^H \big)}^{\substack{ \Mat{I}_{L\,N_{SC}} \\ \text{for OFDM}}} { \Mat{F} \Mat{H}_{\rm eff}} 
  \overbrace{\big( \Mat{I}_L \otimes \Mat{D}_{N_{SC}} \big)}^{\substack{ \Mat{I}_{L\,N_{SC}} \\ \text{for OFDM}}} \Vec{x} + \tilde{\Vec{n}} \nonumber  \\ \notag\\=&   \Mat{K} \Vec{x}  + \tilde{\Vec{n}}  =  \quad \underbrace{\Mat{I}\odot \Mat{K} \Vec{x}}_\text{ desired signal} \quad + \quad  \underbrace{\left(\Mat{K}-\Mat{I}\odot \Mat{K}\right) \Vec{x}}_\text{intra- and interlayer interference} \quad + \quad \tilde{\Vec{n}}~. \label{eq:IO}
\end{align}
\hrulefill

\end{table*}
{LTE} operates on a time-frequency grid as shown in Figure \ref{fig:resource_grid}. The number of subcarriers is always a multiple of twelve; twelve adjacent subcarriers over seven (or six -- in case of extended Cyclic Prefix (CP) successive {OFDM} symbols are called Resource Block (RB). Each {RB} thus consists of $12\times7$ ($12\times6$) Resource Elements (REs), corresponding to the different time-frequency bins.   
A detailed description of {LTE} up- and downlink is available, e.g., in~\cite{Dahlman2011\tempOff{,Cab11}}. 

We focus one those details, necessary to describe our system model at time $n$\footnote{Note that we use the symbol $n$ as time index and the vector $\Vec{n}$ for noise, the distinction should be clear from the context.}. {LTE} employs {OFDM(A)} as physical layer modulation and multiple access scheme in the downlink and {SC-FDM(A)}, i.e., {DFT}-spreaded {OFDM}, in the uplink. In a SC-FDM model, OFDM can be considered a special case. The major difference is an additional spreading and de-spreading stage at the transmitter and receiver, highlighted via dashed boxes in Figure \ref{fig:system_model}.  The common parts of the system model will be described from left to right.

Right after the DFT spreading, the {DMRS} are inserted. {DMRS} will be considered later for the purpose of Channel Estimation (CE). Next, MIMO precoding is carried out, exploiting a set of semi-unitary precoding matrices $\Mat{W}$, pooled in the precoder codebook $\mathcal{W}$, as defined in~\cite{3GPP_TS_36211}. For {LTE-A} uplink transmission, the precoding matrix applied for a given user is equal for all {RB} assigned to this user. In case of spatial multiplexing, each spatial layer is transmitted with equal power. 

Each antenna is equipped with its own {OFDM} modulator, consisting of subcarrier mapping, Inverse Fast Fourier Transform (IFFT) and an {CP} addition. To cope with the channel dispersion and to avoid Inter Symbol Interference (ISI), {LTE} employs a {CP}. As a result of multipath propagation a previous symbol may overlap with the present symbol, introducing {ISI} and impairing the orthogonality between subcarriers, i.e., causing Inter Carrier Interference (ICI)~\cite{Nguyen2002}.  Normal and extended {CP} lengths, with a respective duration of 4.7$\mu$s and 16.7$\mu$s, are standardized, enabling a simple trade-off between {ISI} immunity and {CP} overhead. 

At the transmitter, processing occurs in reversed order. First the OFDM demodulation / FFT takes place to get back into the frequency domain. The immunity to multipath propagation (stemming from the CP) allows to employ one-tap frequency domain equalizers $\Mat{F}$ without performance loss. At last, de-spreading delivers the data estimates.\\

\begin{figure}
\centering

\subfigure[Magnitude values of the T\"oplitz structured MIMO time domain channel matrix with 2 transmit antennas and 4 receive antennas.]{\includegraphics[width=0.35\textwidth]{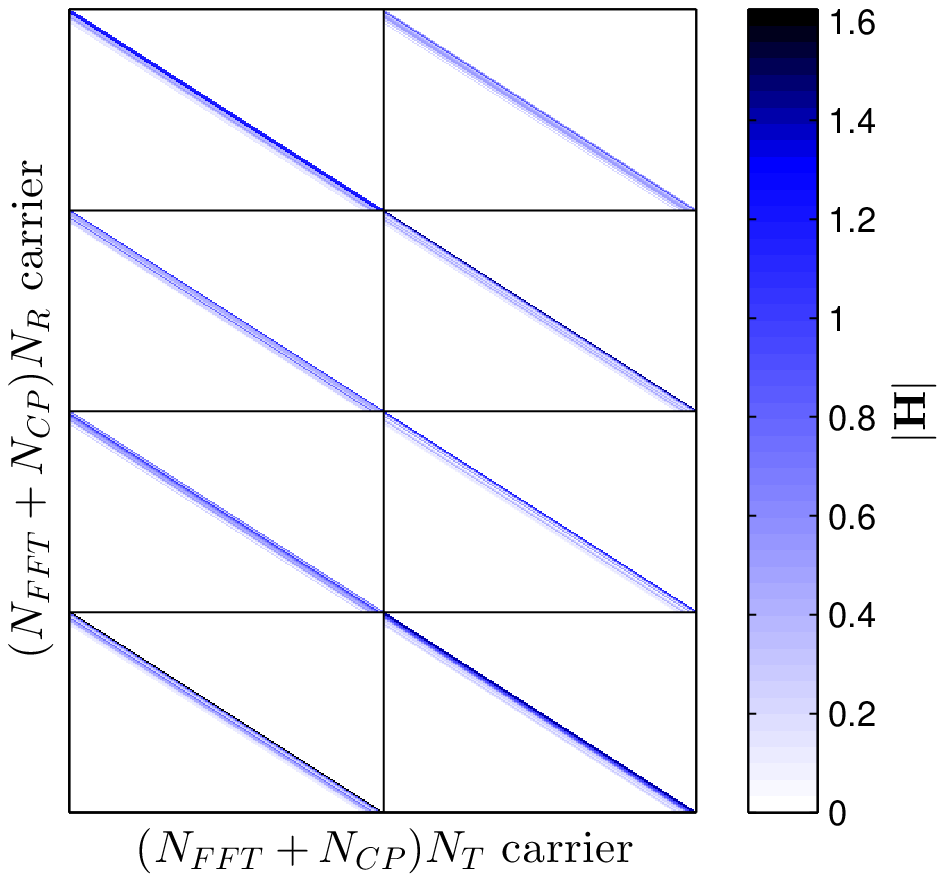}} 


\subfigure[Diagonalized MIMO channel matrix after CP addition/removal and IFFT/FFT at the transmitter and receiver, respectively.]{\includegraphics[width=0.35\textwidth]{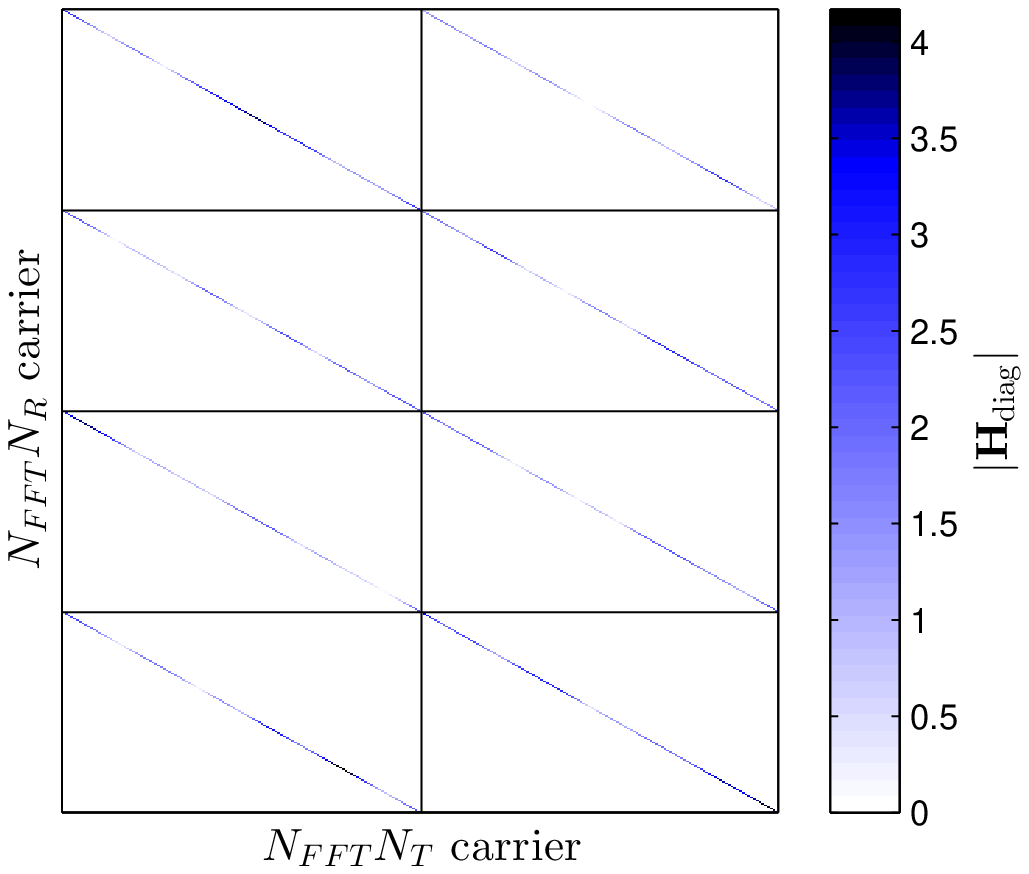}}


\subfigure[Input-Output matrix $\Mat{K}$ employing an MMSE equalizer. The non-diagonal elements represent the intra- and interlayer interference.]{\includegraphics[width=0.35\textwidth]{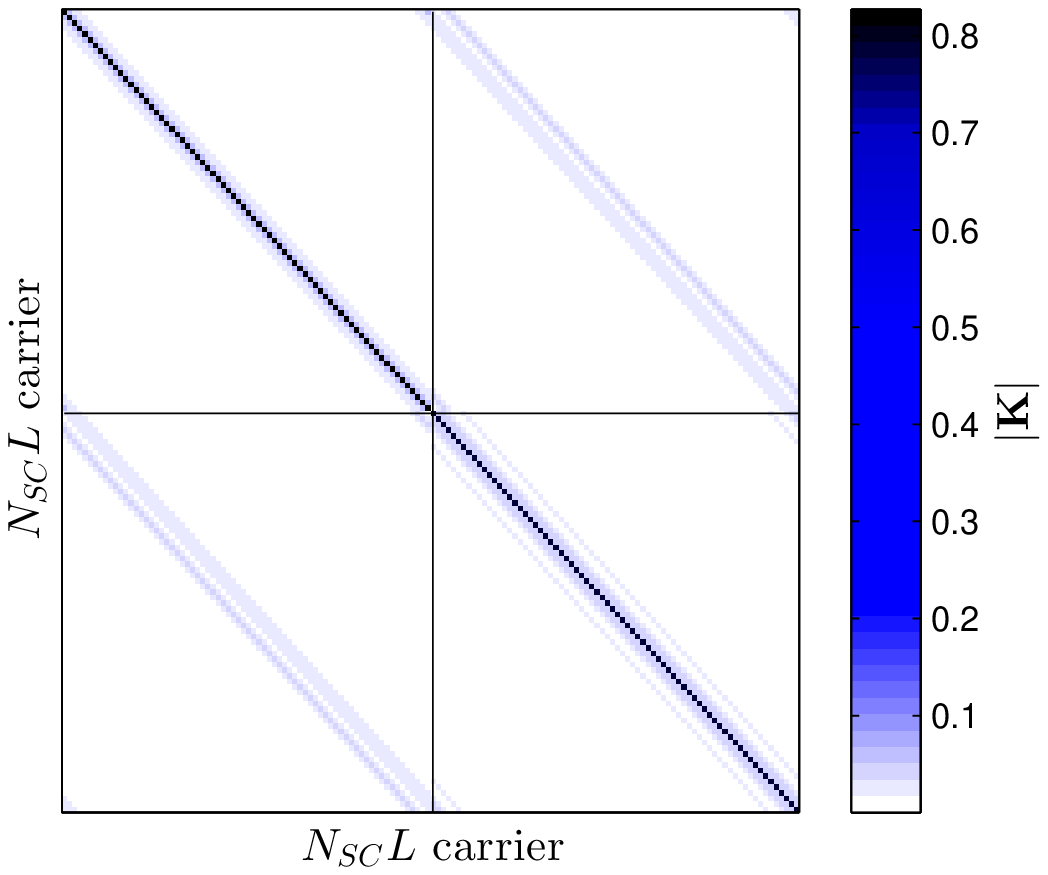}} 


\caption{Examples of different channel abstractions.} \label{fig:channels}
\end{figure}

\noindent All this previously informally described processing is linear and we are able to formulate a matrix-vector input-output relationship between a (stacked) data-vector $\Vec{x}$ and its estimate $\Vec{\hat{x}}$. For simplicity we assume that the channel stays constant during one OFDM symbol. \delete{, and provide {SINR} formulas with and without DFT spreading, i.e., for SC-FDM and OFDM transmission.} A detailed system description based on \cite{wilzeck:model} can be found in \cite{zoechmann:mimotransmission}.

In order to adapt the data transmission to the current channel state, LTE-A applies limited feedback; a  comprehensive 	specification follows in Section \ref{sec:adaptation}.  \delete{divided into the {RI} who chooses the number of spatial layers up to $L=4$ (2 bit), the {PMI} who selects the best fitting spatial precoder out of the codebook $\mathcal{W}$ (5 bit) and the {CQI} who selects  one out of 16 modulation and coding schemes (4 bit). } Limited feedback is depicted via the feedback arrow in Figure \ref{fig:system_model}.
The data vector $\Vec{x}^{(l)} \in\mathbb{C}^{N_{SC}\times 1}$ of layer $l \in\{1,\ldots, L\}$ contains modulated symbols for each of the $N_{SC}$ subcarriers. \delete{As mentioned, t}The number of transmit layers depends on the LTE-A specific Rank Indicator (RI) feedback. The data symbols are coded with a punctured turbo code whose rate is determined by the Channel Quality Indicator (CQI). Subsequently, the codewords are mapped  onto a Quadrature Amplitude Modulation (QAM) alphabet (4/16/64\,{QAM}), where the size of the alphabet depends on the CQI as well.
All $\Vec{x}^{(l)}$ are stacked into one vector $\Vec{x}\in\mathbb{C}^{N_{SC}L\times 1}$ on which layer-wise spreading and joint precoding - according to the Precoding Matrix Indicator (PMI) - of all subcarriers takes place. \delete{Each transmit antenna has its individual} The subsequent OFDM modulator consists of the localized subcarrier mapping $\Mat{M}$, mapping $N_{SC}$ subcarriers to the center of an $N_{FFT}$ point {IFFT}, and the addition of the {CP}. 

Depending on the level of abstraction, our system model can be described via different channel matrices. The physical baseband time domain channel is described  by a block-wise T\"oplitz matrix $\Mat{H} \in \mathbb{C}^{(N_{FFT}+N_{CP})N_R \times (N_{FFT}+N_{CP})N_T}$,  with $N_T$ transmit and $N_R$ receive antennas, which turns block-wise circulant ($\Mat{H}_{\rm cir}$) after addition ($\Mat{P}_{\text{addCP}}$) and removal ($\Mat{P}_{\text{remCP}}$) of an appropriately chosen {CP} of length $N_{CP}$. Finally, it turns diagonal after the IFFT and FFT on the transmitter and receiver, respectively. An example of the T\"oplitz and diagonal structured channel is demonstrated in Figure \ref{fig:channels}~(a)~and~(b), respectively.

\begin{table}[h!]
\begin{equation}
\Mat{H}_{\rm diag} =   \big( \Mat{I}_{N_{R}} \otimes  \Mat{D}_{N_{FFT}} \Mat{P}_{\text{remCP}}\big)  \Mat{H}  
\big( \Mat{I}_{N_{T}} \otimes  \Mat{P}_{\text{addCP}}\Mat{D}_{N_{FFT}}^H \big)
\end{equation}
\end{table}

The last step of the OFDM de-modulator is the reversal of the localized subcarrier mapping $\Mat{M}^H$. The effective MIMO channel from $L$ transmit layers to $N_R$ receive antennas, incorporating the precoder, the OFDM modulator, the time-domain MIMO channel $\Mat{H}$ and the OFDM de-modulator, is abstracted to one block matrix $\Mat{H}_{\rm eff}$. This greatly facilitates the readability of all formulas later on. 
\begin{table}[h!]
\begin{equation}
 \Mat{H}_{\rm eff} =  \big( \Mat{I}_{N_{R}} \otimes  \Mat{M}^H \big)  \Mat{H}_{\rm diag}  
\big( \Mat{I}_{N_{T}} \otimes  \Mat{M}\big)  \big( \Mat{W} \otimes \Mat{I}_{N_{SC}} \big) \label{eq:h_eff}
\end{equation}
\end{table}

The additive noise is assumed independent across antennas and is distributed zero mean, white Gaussian $\Vec{n}^{(i)} \sim \mathcal{CN}\lbrace \Vec{0}, \sigma_n^2 \Mat{I} \rbrace, \;\; i \in \lbrace 1,\dots, N_R\rbrace$. The stacked noise vector $\Vec{n}=\big( (\Vec{n}^{(1)})^T, \dots, (\Vec{n}^{(N_R)})^T  \big)^T$ is thus zero mean, white Gaussian as well.

The frequency domain one-tap equalizer $\Mat{F}$ is chosen conforming to different criteria, either the Zero Forcing (ZF) criterion, which removes all channel distortions at risk of noise enhancement, or the Minimum Mean Squared Error (MMSE) criterion, that tries to minimize the effects of noise enhancement and channel distortion.

After the de-spreading operation the data estimates $\Vec{\hat{x}}$  of the noisy, received signal are given in Equation (\ref{eq:IO}), with the before mentioned convenient  abbreviation (\ref{eq:h_eff}) and $\Mat{D}_{N_{FFT}}$ is the DFT matrix of size $N_{FFT}$. .

\subsection{SC-FDM SINR}

The special structure of Equation (\ref{eq:IO}), due to the frequency domain one tap equalizer and the {DFT} spreading, yields a block-wise circulant input-output matrix, cf. Figure \ref{fig:channels}~(c), 
\be 
\Mat{K} = \big( \Mat{I}_{L} \otimes \Mat{D}_{N_{SC}}^H \big)  \Mat{F} \Mat{H}_{\rm eff}
  \big( \Mat{I}_L \otimes \Mat{D}_{N_{SC}} \big)~.
\ee This block-wise circulant structure produces a constant post equalization and post spreading {SINR} over all subcarriers within one layer \cite{zoechmann:mimotransmission\tempOff{,zoechmann:limitedfeedback}}. The detailed derivation is provided in the appendix.

\begin{eqnarray}
\hspace{-10ex}&{\operatorname{{SINR}}}^{{\rm SC-FDM},\;(l)} =  \label{eq:SNR_SCFDMA} \\
\hspace{-10ex}&\frac{\frac{\sigma_x^2}{N_{SC}}\big| \mathds{1}_{_{N_{SC}}}^T \Mat{S}^{(l)} \diag \!\!\!\left( \Mat{F}\Mat{H}_{\rm eff} \right)\! \big|^2}{\!\!\sigma_x^2 \| \Mat{S}^{(l)} \Mat{F}\Mat{H}_{\rm eff} \|_F^2  -  \frac{\sigma_x^2}{N_{SC}}\big| \mathds{1}_{_{N_{SC}}}^T \Mat{S}^{(l)} \diag \!\!\!\left( \Mat{F}\Mat{H}_{\rm eff} \right)\! \big|^2 +\sigma^2_n \| \Mat{S}^{(l)} \Mat{F} \|_F^2 } \nonumber ,
\end{eqnarray}
where 
\be
\label{eq:selector_matrix}
\Mat{S}^{(l)}=\left(\Mat{0} \quad \Mat{I}_{N_{SC}} \quad \Mat{0}  \right)~,
\ee
selects that part of $\Mat{F}\Mat{H}_{\rm eff}$ effecting the $l^{\rm th}$ layer. The second moment of the zero mean symbols $\sigma_x^2$ equals the baseband transmit power as {LTE-A} has standardized semi-unitary precoders $W$, so that the overall transmitter (spreading, precoding and OFDM modulation) is unitary.

\subsection{OFDM SINR}

In contrast to {SC-FDM}, no spreading takes place for OFDM. The dashed boxes in Figure \ref{fig:system_model} are replaced by identity matrices; they are simply omitted. Thus, different subcarriers $k$ are orthogonal/independent and the equalizer treats the corresponding subcarrier channel $\Mat{H}_{k}$ only. We use the subscript $k$ to denote the relevant part of the full channel matrix $\Mat{H}_{\rm eff}$ for the $k^{\rm th}$ subcarrier.  The corresponding indices within the diagonal matrix $\Mat{H}_{\rm diag}$ are $\mathds{1}_{N_{R}\times N_T} \otimes \mathrm{Diag}\left(\Vec{e}_k\right)$, with the canonical base vectors $\Vec{e}_k$. Using this notation, the effective subcarrier channel $\Mat{H}_{k} \in \mathbb{C}^{N_{R}\times L}$ is 
\be
\Mat{H}_{k} = \left[\Mat{H}_{\rm diag}\right]_{\mathds{1}_{N_{R}\times N_T} \otimes \mathrm{Diag}\left(\Vec{e}_k\right)} \Mat{W}~,
\ee
and $\Mat{F}_k$ is its linear one tap equalizer.
The {SINR} formula is quite similar to the {SC-FDM} case, except that the {SINR} shows subcarrier dependency now. The {SINR} vector at layer $l$ reads
\begin{eqnarray}
\hspace{-10ex}&\left[\Vec{\operatorname{\mathbf{SINR}}}^{{\rm OFDM},\;{(l)}}\right]_k= \label{eq:SNR_OFDMA} 
\\ \hspace{-10ex}&\frac{\sigma_x^2 \big|\Mat{s}^{(l)}\diag \!\!\!\left( \Mat{F}_k \Mat{H}_{k} \right)\! \big|^2 }{ { \sigma_x^2 \| \Mat{s}^{(l)} \Mat{F}_k\Mat{H}_{k} \|_2^2-\sigma_x^2 \big|\Mat{s}^{(l)} \diag \!\!\!\left( \Mat{F}_k \Mat{H}_{k}  \right)\!\big|^2 }+\sigma^2_n \| \Mat{s}^{(l)} \Mat{F}_k \|_2^2 } ~, \notag
\end{eqnarray} 
with the selection vector 
\be 
\Mat{s}^{(l)} = \left( \!\begin{array}{ccc}
{0} \dots 0 & {1} & 0 \dots {0}
\end{array}\!  \right) ~,
\ee
with appropriate number of zeros and a one at the $l^{\rm th }$ position.


\section{SC-FDM Effects}
\label{sec:effects}

\subsection{Peak to Average Power Ratio}

{SC-FDM} is employed as the physical layer modulation scheme for {LTE} uplink transmission, due to its lower {PAPR} compared to {OFDM}~\cite{myung2006single}. Lower {PAPR}, or similarly lower crest factor, leads to reduced linearity requirements for the power amplifiers and to relaxed resolution specifications for the digital-to-analog converters at the user equipments, entailing higher power efficiency. 

The Vienna LTE-A uplink simulator calculates the discrete-time baseband {PAPR} with the default oversampling factor $J=4$~\cite{jiang2008overview}.
The discrete time signal on transmit antenna $t \in \lbrace 1,\dots,N_T\rbrace$ is therefore calculated as
\begin{eqnarray}
\hspace{-1ex}\left[\Vec{s}_{\rm tx}^{(t)}\right]_m\!\!=\!\!\frac{1}{\sqrt{N_{_{FFT}}}} \hspace{-2ex}\sum_{k=0}^{N_{FFT}-1} \hspace{-2ex}\left[\Vec{x}_{\rm pre}^{(t)}\right]_k e^{j\frac{2\pi m k}{J N_{FFT}}}, \\ \;\; 0 \le m \le J N_{_{FFT}}-1 ~, \notag
\end{eqnarray}
where $\Vec{x}_{\rm pre}^{(t)}$ is the transmit vector right after precoding and before the IFFT at transmit antenna $t$. The {PAPR} of the stacked vector $\Vec{s}_{\rm tx}=\big( (\Vec{s}_{\rm tx}^{(1)})^T, \dots, (\Vec{s}_{\rm tx}^{(N_T)})^T  \big)^T$ is calculated as 
\begin{eqnarray}
\operatorname{PAPR}\{\Vec{s}_{\rm tx}\}&= \frac{\max\limits_{1\le t \le N_T} \max\limits_{0 \le m \le J N_{_{FFT}}-1}\left(\big|\big[\Vec{s}_{\rm tx}^{(t)}\big]_m \big|^2  \right)}{\mathbb{E}_t\left\{\mathbb{E}_n\big\lbrace\big|\big[\Vec{s}_{\rm tx}^{(t)}\big]_m \big|^2 \big\rbrace\right\}} \\
&\approx N_T N_{_{FFT}} \| \diag\left(\Vec{s}_{\rm tx} \Vec{s}_{\rm tx}^H  \right)\|_\infty \big/ \|\Vec{s}_{\rm tx} \|_2^{2}~,\notag
\end{eqnarray}
where the Euclidean norm in the denominator serves as an estimate for the ensemble average.
\begin{figure}
\centering
\includegraphics[width=0.45\textwidth]{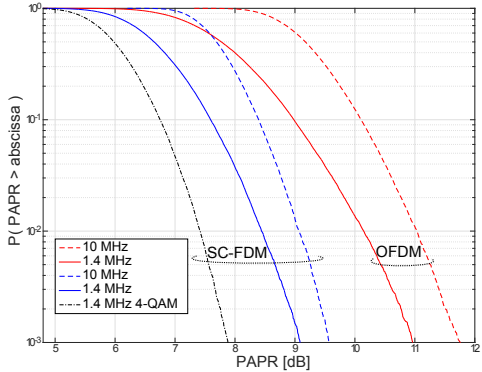}
\caption{PAPR for SC-FDM and OFDM for different bandwidths  (1.4 MHz and 10 MHz) and modulation alphabets (4/64 QAM).} \label{fig:papr}
\end{figure}

Figure \ref{fig:papr} depicts the {PAPR} of {OFDM} and {SC-FDM} obtained for different system bandwidths. Already for a small bandwidth (1.4 MHz), there is a significant reduction for {SC-FDM} over {OFDM}. With increasing bandwidth OFDM's {PAPR} grows and the gains obtained by SC-FDM become more and more pronounced. The {PAPR} also depends on the modulation alphabet; the smaller the alphabet, the smaller the {PAPR}. This effect is illustrated in dotted lines in Figure \ref{fig:papr}, where we have shown the PAPR of 4-QAM, exemplarily.

\subsection{BER Comparison over Frequency Selective Channels}

The additional spreading of SC-FDM leads to an SINR expression that is constant on all subcarriers as for single carrier transmission, legitimating its name. The aim of this subsection is to analyse the SINR expression more carefully for the Single Input Single Output (SISO) case\footnote{The reduction to SISO is done to make our results comparable even to older Frequency Domain Equalization (FDE) works, e.g., \cite{shi2004capacity}} and draw conclusions on {BER} performance.
\begin{figure}
\centering

\subfigure[ZF receiver]{\includegraphics[width=0.4\textwidth]{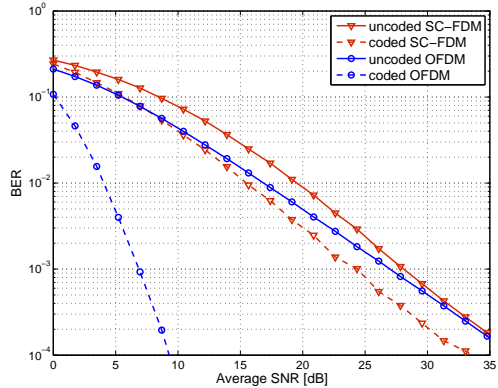}} 


\subfigure[MMSE receiver]{\includegraphics[width=0.4\textwidth]{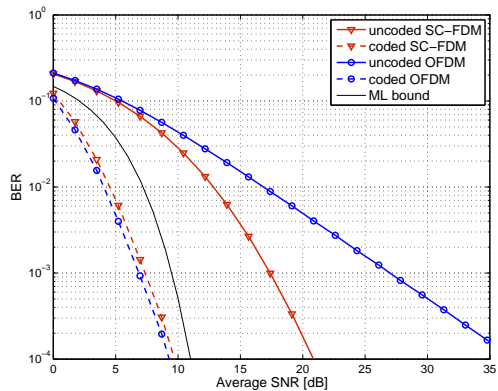}} 


\caption{BER comparison between OFDM and SC-FDM for a SISO PedB channel with 5 MHz bandwidth and fixed ${\rm CQI} = 4$ transmission.}\label{fig:BER}

\end{figure}

We focus on the two most prominent equalizer concepts and start with the ZF equalizer, for whom the {SC-FDM} Signal to Noise Ratio (SNR) expression (\ref{eq:SNR_SCFDMA}) reduces to the harmonic mean
\be 
{\operatorname{SNR}}_{\rm ZF}^{\rm SC-FDM}=\frac{\sigma_x^2}{\sigma^2_n}  \frac{1}{\frac{1}{N_{SC}}\sum\limits_{k=1}^{N_{SC}} \frac{1}{|\Mat{H}_k|^2}} ~, \label{eq:harmonic_SNR}
\ee
whereas the {OFDM} expression (\ref{eq:SNR_OFDMA}) is sub-carrier dependent and becomes proportional to the channel transfer function
\be 
\left[\Vec{\operatorname{SNR}}_{\rm ZF}^{\rm OFDM}\right]_k=\frac{\sigma_x^2}{\sigma^2_n}  |\Mat{H}_k|^2 ~.
\ee
The average {OFDM} {SNR} 
\be  \label{eq:avag_OFDM}
\overline{ {\operatorname{SNR}}_{\rm ZF}^{\rm OFDM}}=\frac{\sigma_x^2}{\sigma^2_n} \frac{1}{N_{SC}}\sum\limits_{k=1}^{N_{SC}} |\Mat{H}_k|^2 
\ee
yields an upper bound on the Single Carrier Frequency Division Multiple Access (SC-FDMA) {SNR} due to the harmonic mean -- arithmetic mean inequality~\cite{bullen2003handbook}.
\be  \label{eq:bound_ZF}
{\operatorname{SNR}}_{\rm ZF}^{\rm SC-FDM} \le \overline{ {\operatorname{SNR}}_{\rm ZF}^{\rm OFDM}}
\ee
Equality in Equation (\ref{eq:bound_ZF}) holds if and only if the channel is frequency flat. The difference between the harmonic mean and the arithmetic mean gets increasingly pronounced, the more selective the channel becomes. We therefore expect the (uncoded) {BER} of SC-FDM and ZF equalization to perform worse than OFDM, which is also validated by simulations. The BER simulations were carried out with ${\rm CQI} = 4$ on a PedB channel \cite{tr25943}. This Modulation and Coding Scheme (MCS) employs 4-{QAM} and has an effective code-rate of $0.3008$. As expected, the BER performance of SC-FDM is worse than OFDM, both shown in Figure \ref{fig:BER}~(a) in solid lines. Due to the spreading SC-FDM already expends all channel diversity and coding does not increase the SNR slope of the BER curve. This manifests in an almost parallel shift of the BER curve for SC-FDM, as visual in Figure \ref{fig:BER}~(a) in red dashed lines. None exploited diversity allows OFDM to increase the BER slope considerably, cf. Figure \ref{fig:BER}~(a) blue dashed line. \tempOff{Closed form solutions for the {BER} achieved with {ZF} receivers in {SC-FDM} transmission are given in~\cite{BER_SCFDMA}.}

The {MMSE} SINR expression is less intuitive and for the purpose of comparison, similar mathematical transformations as in \cite{nisar2007performance}  and \cite{sanchez2011ber} are  required to arrive at
\begin{eqnarray} \label{eq:SISO_SC_SNR}
{\rm SINR}_{\rm MMSE}^{\rm SC-FDM} &= \frac{\sigma_x^2}{\sigma_n^2}     \frac{1-\frac{\sigma_n^2}{\sigma_x^2} \frac{1}{N_{SC}} \sum\limits_{k=1}^{N_{SC}} \frac{1}{\frac{\sigma_n^2}{\sigma_x^2}+|\Mat{H}_k|^2} }{\frac{1}{N_{SC}}\sum\limits_{k=1}^{N_{SC}} \frac{1}{\frac{\sigma_n^2}{\sigma_x^2}+|\Mat{H}_k|^2}} \\ &=\frac{\sigma_x^2}{\sigma_n^2}\bigg(\frac{1}{\frac{1}{N_{SC}}\sum\limits_{k=1}^{N_{SC}} \frac{1}{\frac{\sigma_n^2}{\sigma_x^2}+|\Mat{H}_k|^2}} -\frac{\sigma_n^2}{\sigma_x^2} \bigg)~. \notag
\end{eqnarray} 
The detailed derivation is shown in the appendix. The denominator of Equation (\ref{eq:SISO_SC_SNR}) is regularized and less sensitive to spectral notches.

An upper bound on the SINR can be obtained via the maximum of the transfer function $\Mat{H}_k$
\begin{eqnarray}
{\rm SINR}_{\rm MMSE}^{\rm SC-FDM} &\le \frac{\sigma_x^2}{\sigma_n^2}\bigg(\frac{1}{ \frac{1}{ \;\; \frac{\sigma_n^2}{\sigma_x^2}+\max_k|\Mat{H}_k|^2} \;\;} -\frac{\sigma_n^2}{\sigma_x^2} \bigg) \notag \\ &= \frac{\sigma_x^2}{\sigma_n^2} \max_k|\Mat{H}_k|^2 \label{eq:SNR_bound} ~.
\end{eqnarray}
In the low SNR regime $\frac{\sigma_n^2}{\sigma_x^2} \gg |\Mat{H}_k|^2$ this bound becomes tight. The higher the inverse SNR $\frac{\sigma_n^2}{\sigma_x^2}$ in relation to the maximum of the transfer function, the tighter the bound becomes. As Equation (\ref{eq:avag_OFDM}) -- the average OFDM SNR -- can never be larger to its maximum entry (only equal for frequency flat channels), the SC-FDM SINR expression (\ref{eq:SISO_SC_SNR}) is larger than (\ref{eq:avag_OFDM})\footnote{For the SISO case, ZF and MMSE equalizers perform equivalent for OFDM and 4-{QAM}.} at low SNR, as predicted by the tight bound  (\ref{eq:SNR_bound}); a lower BER is expected. Again, this presumption is validated by our simulation, showing that the uncoded BER is lower for SC-FDM as for ODFM, cf. Figure \ref{fig:BER}~(b) in solid lines. \tempOff{There are also semi-closed form approximations for the {BER} of {MMSE} detection in {SC-FDM} transmission available~\cite{sanchez2011ber}.} Although the uncoded BER shows superior performance, the coded BER is lower for OFDM due to the coding gains stemming from channel diversity, cf., Figure \ref{fig:BER}~(b) dashed lines. 

A bound for the Maximum Likelihood (ML) detection performance was derived in~\cite{Geles_ML_bound}. As bandwidth increases the slope of the {BER} curve achieved with {MMSE} receivers tends to the slope of {ML} detection, demonstrating the full exploitation of channel diversity by the {MMSE} equalizer, cf., Figure \ref{fig:BER}~(b) black line.




\section{Link Adaptation}
\label{sec:adaptation}

\begin{figure*}
\centering
%
	\subfigure[ZF receivers\label{fig:TP_2x2_ZF}] {\includegraphics[width=0.45\textwidth]{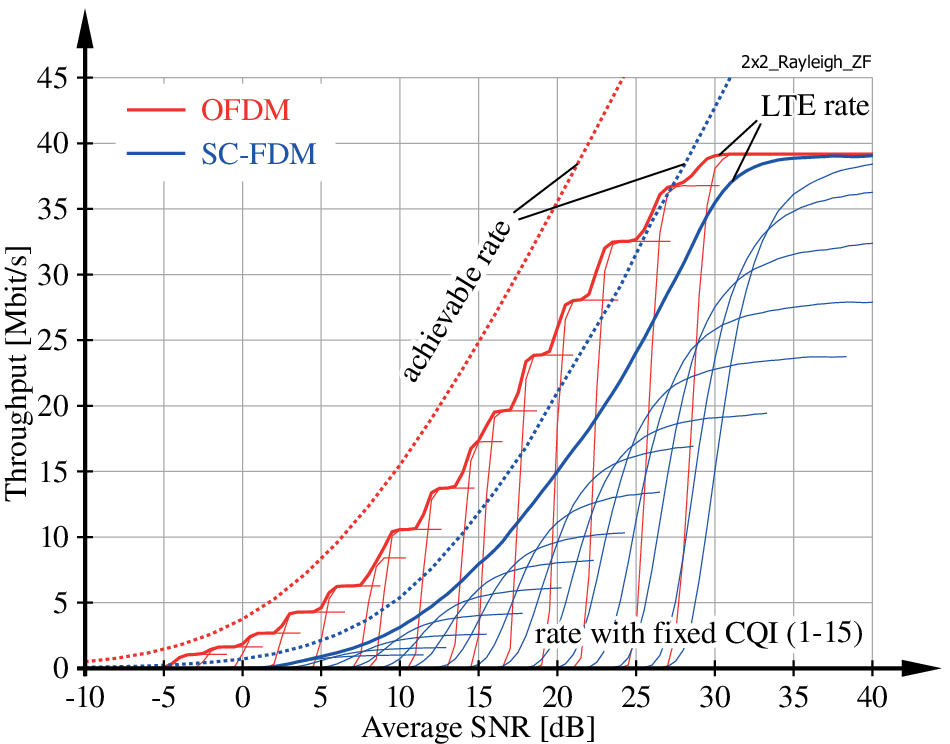}}\qquad
	\subfigure[MMSE receivers]{\includegraphics[width=0.45\textwidth]{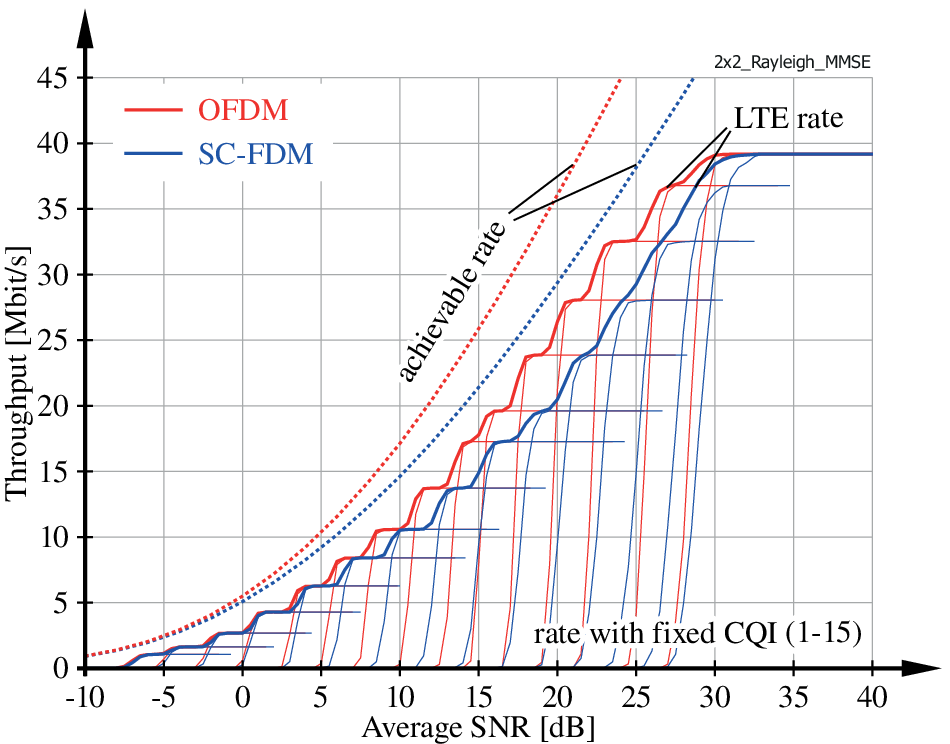}}
	\caption{Throughput comparison of OFDM and SC-FDM with rate adaptation and $2\times 2$ Rayleigh fading channels of 5\,MHz bandwidth.}
	\label{fig:TP_2x2}
\end{figure*}

In this section, we first investigate the throughput performance of {LTE-A} uplink employing ideal rate adaptation and compare {SC-FDM} transmission to {OFDM} with {ZF} and {MMSE} receivers. Then, we extend our single-user {MIMO} {CSI} feedback algorithms proposed for {LTE} downlink in~\cite{Sch11} to {LTE} uplink and evaluate their performance comparing to the throughput bounds developed in~\cite{Schwarz-bounds2011}. We also highlight some important basic differences between link adaptation in {LTE} up- and downlink transmissions.     

\subsection{Performance with Ideal Rate Adaptation}

As demonstrated in the previous section, {SC-FDM} provides a significant advantage in terms of {PAPR} over {OFDM}, thus relaxing linearity requirements of radio frequency power amplifiers for user equipments. Yet, this comes at the cost of coded {BER} degradation since channel diversity is lost and the performance is mostly dominated by the weakest subcarrier of a user, especially with {ZF} receivers; c.f.,~(\ref{eq:harmonic_SNR}). This diversity loss cannot be recovered by forward-error-correction channel coding, since the {DFT}-spreading applied with {SC-FDM} effectively causes an averaging over {SINR} observed on all scheduled subcarriers according to~(\ref{eq:SNR_SCFDMA}). As a consequence, {SC-FDM} transmission over frequency selective channels achieves worse throughput than {OFDM}. This is demonstrated in Figure~\ref{fig:TP_2x2}, where we cross-compare the achievable rate, as defined in Equation~(\ref{eq:OF_rate}) and (\ref{eq:SC_rate}), and the actual throughput of {SC-FDM} and {OFDM} transmission as obtained by the Vienna {LTE}-A uplink simulator. We consider single-user transmission over $5\,$MHz bandwidth assuming $N_T = N_R = 2$ antennas at the user and the base station and $L = 2$ spatial layers. The precoder is selected as a scaled identity matrix: $\Mat{W} = 1/\sqrt{L}\, \Mat{I}_L$. We consider transmission over independent and identically distributed frequency-selective Rayleigh fading channels, emphasizing the difference between {OFDM} and {SC-FDM}. The achievable rate in bits per {OFDM}/{SC-FDM} symbol with Gaussian signalling and equal power allocation over subcarriers and spatial layers is calculated as
\begin{gather}
	R^{\rm OFDM} = \sum_{k = 1}^{N_\text{SC}} \sum_{l = 1}^{L} \mathrm{log}_2\left(1 + \left[\Vec{\operatorname{{SINR}}}^{{\rm OFDM},\;(l)}\right]_k \right), \label{eq:OF_rate} \\
	R^{\rm SC-FDM} = N_\text{SC} \sum_{l = 1}^{L} \mathrm{log}_2\left(1 + {\operatorname{{SINR}}}^{{\rm SC-FDM},\;(l)} \right), \label{eq:SC_rate} 
\end{gather}      
with the receiver-specific post-de-spreading (post-equalization) {SINR}s from~(\ref{eq:SNR_SCFDMA}) and~(\ref{eq:SNR_OFDMA}), respectively.   

We observe a significant loss of achievable rate of {SC-FDM} transmission compared to {OFDM} in Figure~\ref{fig:TP_2x2}, which is especially pronounced with {ZF} receivers due to noise enhancement. In Figure~\ref{fig:TP_2x2}, we also show the actual rate achieved by {LTE} uplink {SC-FDM} transmission with ideal rate adaptation and compare to the performance obtained by {OFDM} transmission; the corresponding curves are denoted by \emph{{LTE} rate}. We determine the performance of ideal rate adaptation by simulating all possible transmission rates, corresponding to {CQI}1 to {CQI}15, and selecting at each subframe the largest rate that achieves error free transmission. The figure also shows the throughput of the individual {CQI}s. We observe a gap between the {LTE} throughput with {OFDM} and {SC-FDM} that is similar to the gap in terms of achievable rate. Notice that the performance loss with {MMSE} receivers is significantly smaller than with {ZF} detection, since {MMSE} avoids excessive noise enhancement. 

We also observe in Figure~\ref{fig:TP_2x2_ZF} that the gain achieved by instantaneous rate adaptation, as compared to rate adaptation based on the long-term average {SNR}, is much larger for {ZF} {SC-FDM} than for {ZF} {OFDM}; this is evident from the distance between the curves with rate adaptation (\emph{LTE rate}) and the curves with \emph{fixed {CQI}}. The reason for this behaviour is that the {SNR} of {ZF} {SC-FDM} shows strong variability around its means, since it is dominated by the worst-case per-subcarrier {SNR} according to~(\ref{eq:harmonic_SNR}); the average {SNR} over subcarriers of {ZF} {OFDM}, however, approximately coincides with its mean value. This implies that the optimal {CQI} of {ZF} {SC-FDM} can vary significantly in-between subframes, as reflected by the large average {SNR} variation required to increase the rate with fixed {CQI} from zero to its respective maximum.  Yet, for {ZF} {OFDM} the throughput of the individual {CQI}s follows almost a step function; hence, rate adaptation can be based on the long-term average {SNR} without substantial performance degradation.\footnote{Notice, however, that instantaneous rate adaptation for {ZF} {OFDM} can be advantageous in case of frequency-correlated channels\cite{Schwarz-WiAd2010}.}          

\begin{figure*}
\centering
	\subfigure[Comparison of achievable rate and the rate estimate of Equation~(\ref{eq:rate_est}).]{\includegraphics[width=0.45\textwidth]{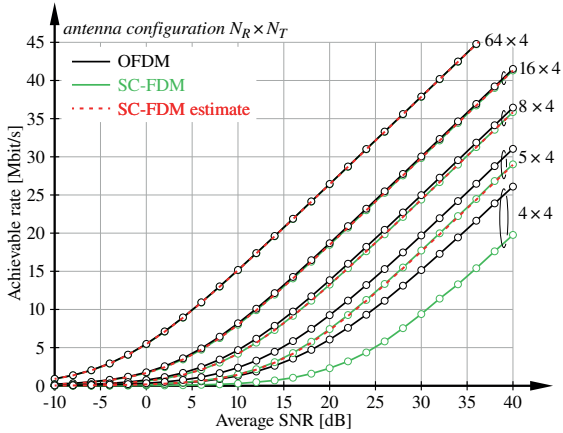}}\qquad
	\subfigure[Relative achievable rate of {SC-FDM} with respect to {OFDM}.]{\includegraphics[width=0.45\textwidth]{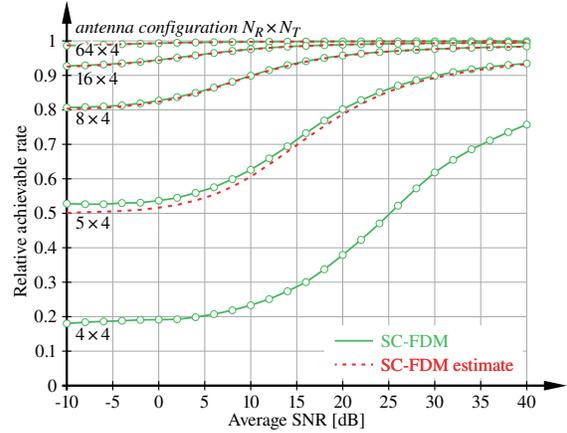}}
	\caption{Achievable rate of {OFDM} and {SC-FDM} with {ZF} equalizers and growing number of receive antennas at fixed number of streams $L = 4$.}
	\label{fig:TP_Xx4}
\end{figure*}	

In case $N_R > L$, we can easily estimate the achievable rate of {SC-FDM} transmission: The per-layer {SNR} with {ZF} receivers is governed by the harmonic mean of the channel responses on the individual subcarriers, similar to~(\ref{eq:harmonic_SNR})
\begin{gather}
	\hspace{-8ex} \operatorname{{SNR}}_{\rm ZF}^{{\rm SC-FDM},\;(l)}\hskip-3pt =\hskip-2pt \frac{\sigma_x^2/\sigma_n^2}{\frac{1}{N_\text{SC}} \sum_{k = 1}^{N_\text{SC}}\hskip-2pt \left[\left(\left(\Mat{H}_k \Mat{W}\right)^\mathrm{H} \left(\Mat{H}_k \Mat{W}\right)\right)^{-1}\right]_{l,l}}, \label{eq:SINR_MIMO_ZF}
\end{gather}    
with $\Mat{H}_k \in \mathbb{C}^{N_R \times N_T}$ denoting the {OFDM} channel matrix on subcarrier $k$. Assuming constant precoding and semi-correlated Rayleigh fading 
\begin{gather}
	\Mat{H}_k = \tilde{\Mat{H}}_k \Mat{C}_T^{\frac12},\quad \left[\tilde{\Mat{H}}_k\right]_{i,j} \sim \mathcal{CN}\left\{0,1\right\},
\end{gather}
with $\Mat{C}_T \in \mathbb{C}^{N_T \times N_T}$ determining the spatial correlation at the user equipment side, the matrix in the denominator of~(\ref{eq:SINR_MIMO_ZF}) follows a complex inverse Wishart distribution with $N_R$ degrees of freedom and scale matrix $\Mat{C} = \left(\Mat{W}^\mathrm{H} \Mat{C}_T \Mat{W}\right)^{-1}$
\begin{gather}
	\overline{\Mat{H}} = \left(\left(\Mat{H}_k \Mat{W}\right)^\mathrm{H} \left(\Mat{H}_k \Mat{W}\right)\right)^{-1} \sim \mathcal{CW}_L^{-1}\left\{N_R,\Mat{C}\right\}.
\end{gather}
Letting $N_\text{SC} \rightarrow \infty$, we can replace the term in the denominator of~(\ref{eq:SINR_MIMO_ZF}) with its expected value
\begin{gather}
	\frac{1}{N_\text{SC}} \sum_{k = 1}^{N_\text{SC}} \left[\;\overline{\Mat{H}}\;\right]_{l,l} \overset{N_\text{SC} \rightarrow \infty}{\longrightarrow} \mathbb{E}\left(\left[\;\overline{\Mat{H}}\;\right]_{l,l} \right).
\end{gather} 
This expected value only exists in case $N_R > L$~\cite{Maiwald2000}. For $N_R = L$, the diagonal elements of $\overline{\Mat{H}}$ follow a heavy-tailed inverted Gamma distribution~\cite{Gupta2000,Rongtao2009} with non-finite first moment. Yet, for $N_R > L$, which is a common situation in cellular networks since the base station is mostly equipped with far more antennas than the users, the expected value is
\begin{gather}
	\mathbb{E}\left(\left[\;\overline{\Mat{H}}\;\right]_{l,l} \right) = \frac{1}{N_R - L} \left[\Mat{C}\right]_{l,l}.
\end{gather} 
Hence, we can estimate the achievable rate of {SC-FDMA} transmission over semi-correlated Rayleigh fading channels
\begin{gather}
	R^{\rm SC-FDM} \approx N_\text{SC} \sum_{l = 1}^{L} \mathrm{log}_2\left(1 + \frac{\sigma_x^2/\sigma_n^2}{\left[\Mat{C}\right]_{l,l}} (N_R-L) \right) \label{eq:rate_est}\\
	 \approx N_\text{SC} L \left[\mathrm{log}_2\left(\frac{\sigma_x^2/\sigma_n^2}{\left(\prod_{l = 1}^L \left[\Mat{C}\right]_{l,l}\right)^{1/L}} \right) + \mathrm{log}_2\left(N_R-L\right)\right]. \label{eq:high_SNR}
\end{gather} 
Here~(\ref{eq:high_SNR}) resembles the high {SNR} approximation of the achievable rate of {OFDM} transmission with {ZF} detection as proposed in~\cite[Eq.\,(14)]{Gore2002}; even more, for fixed $L$ and letting $N_R$ grow to infinity,~(\ref{eq:high_SNR}) and~\cite[Eq.\,(14)]{Gore2002} tend to the same limit, due to channel hardening on each subcarrier with growing number of receive antennas.

In Figure~\ref{fig:TP_Xx4}, we investigate the performance of the rate estimate~(\ref{eq:rate_est}) for $N_T = L = 4$ and varying number of receive antennas. We assume $\Mat{W} = 1/\sqrt{L}\, \Mat{I}_L$ and 
\begin{gather}
	\Mat{C}_T = \left[\begin{array}{cccc} 1 & 0.9 & \ldots & 0.9 \\ 0.9 & \ddots &  & \vdots  \\ \vdots & & & 0.9 \\ 0.9 & \ldots & 0.9 & 1\end{array}\right], \nonumber
\end{gather}	
and consider the smallest {LTE} bandwidth of $N_{\text{SC}} = 72$ subcarriers. We observe that the proposed estimate performs very well even at this small bandwidth; notice, though, that a more realistic channel model with correlation over subcarriers may require larger bandwidth to validate the proposed estimate. Figure~\ref{fig:TP_Xx4} also confirms the observation that single-user {MIMO} {OFDM} and {SC-FDM} with {ZF} detectors tend to the same limiting performance with increasing number of receive antennas. 

This statement, however, will not hold true if the total number of layers grows proportionally with the number of receive antennas. For example, multi-user {MIMO} transmission with {ZF} equalization and single antenna users achieves only a diversity order of $N_R - L +1$~\cite{Hedayat2007}, with $L$ denoting the total number of layers being equal to the number of spatially multiplexed users. Hence, if $L$ scales proportionally with $N_R$, channel hardening on each subcarrier will not occur and thus the performance of {OFDM} and {SC-FDM} will not coincide.   


\subsection{Performance with Realistic Link Adaptation}

\begin{figure*}
\centering
%
	\subfigure[Absolute throughput]{\includegraphics[width=0.45\textwidth]{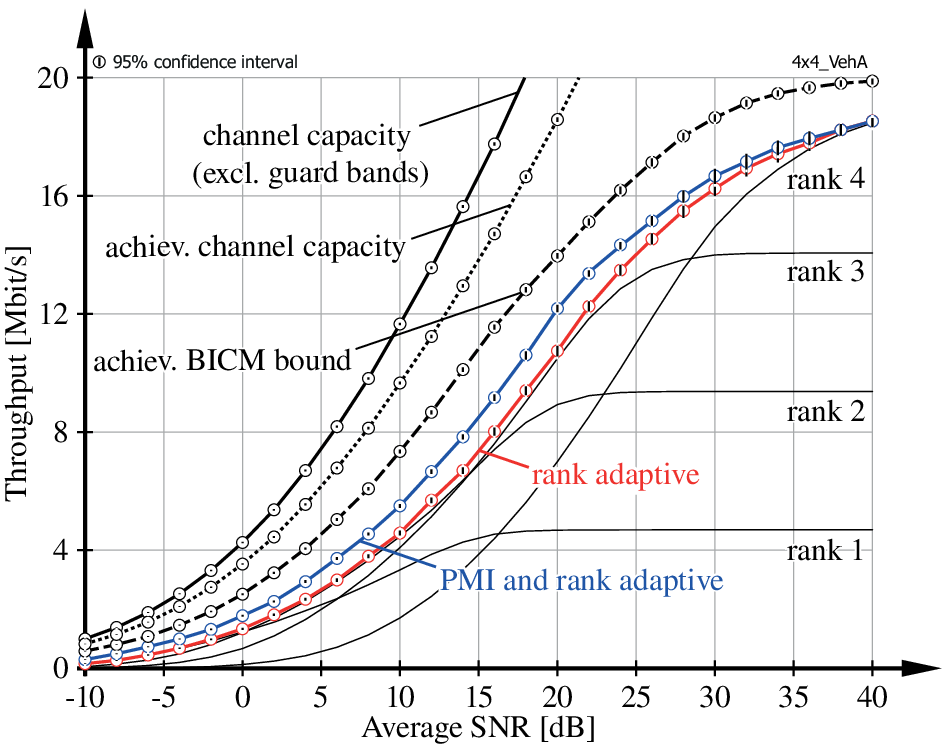}}\qquad
	\subfigure[Relative throughput]{\includegraphics[width=0.45\textwidth]{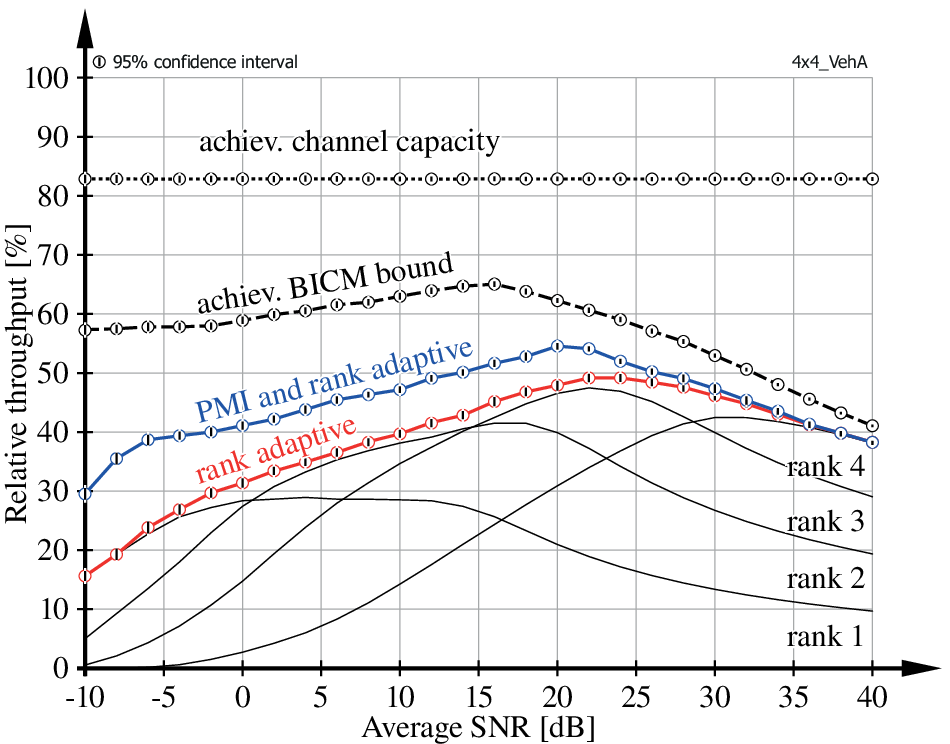}}
	\caption{Absolute and relative throughput of LTE uplink transmission over $4\times 4$ VehA channels of 1.4\,MHz bandwidth employing rate adaptation. We compare the performance of fixed rank, rank adaptive and PMI $+$ rank adaptive transmission to the performance bounds proposed in~\cite{Schwarz-bounds2011}.}
	\label{fig:TP_4x4}
\end{figure*}

Instantaneous rate adaptation is an important tool for exploiting diversity of the wireless channel in {LTE}, by adjusting the transmission rate according to the current channel quality experienced by a user. {LTE} specifies a set of fifteen different {MCS}s; the selected {MCS} is signalled by the {CQI}.

{LTE} additionally supports spatial link adaptation by means of codebook based precoding with variable transmission rank. With this method, the precoding matrix $\Mat{W} \in \mathbb{C}^{N_T \times L}$ satisfying $\Mat{W}^\mathrm{H} \Mat{W} = 1/L\, \Mat{I}_L$ is selected from a standard defined codebook $\mathcal{W}_L$ of scaled semi-unitary matrices; furthermore, the number of spatial layers $L$ can be adjusted to achieve a favourable trade-off between beamforming and spatial multiplexing. The selected precoder and transmission rank are signalled, employing the {PMI} and the {RI}. In single-user {MIMO} {LTE} uplink transmission, the same precoder is applied on all {RB}s that are assigned to a specific user, whereas frequency-selective precoding is supported in {LTE} downlink. 

There is a basic difference between the utilization of {CQI}, {PMI} and {RI} in up- and downlink directions of Frequency Division Duplex (FDD) systems. In downlink, the base station is reliant on {CSI} feedback from the users for link adaptation and multi-user scheduling~\cite{Schwarz-Asilomar2010}, since channel reciprocity cannot be exploited in {FDD}. {CQI}, {PMI} and {RI} can be employed to convey such {CSI} from the users to the base station via dedicated feedback channels~\cite{Schwarz-WiAd2010}. In the uplink, on the other hand, the base station can by itself determine {CSI} exploiting the Sounding Reference Signals (SRSs) transmitted by the users. In this case, {CQI}, {PMI} and {RI} are employed by the base station to convey to the users its decision on link adaptation that has to be applied by the users during uplink transmission. 

In principal, link adaptation must be jointly optimized with multi-user scheduling to optimize the performance of the system, since the effective {SC-FDM} {SINR} (and thus the rate) of a user depends on the assigned {RB}s according to~(\ref{eq:SNR_SCFDMA}). For reasons of computational complexity, however, we assume that the multi-user schedule is already fixed and determine link adaptation parameters based on this resource allocation. 
We modify the approach proposed in~\cite{Sch11} for {LTE} downlink transmission to determine the link adaptation parameters in four steps:

\begin{enumerate}[leftmargin=*]
	\item Determine the optimal precoder for each transmission rank $L \leq \min\left(N_T,N_R\right)$ by maximizing transmission rate
	\begin{gather}
		\hat{\Mat{W}}(L) = \argmax_{\Mat{W} \in \mathcal{W}_L} \sum_{l = 1}^{L} f\left({\operatorname{{SINR}}}^{{\rm SC-FDM},\;(l)} \left( \Mat{W} \right)\right).
	\end{gather}
	Here, function $f(\cdot)$ maps $\text{SINR}$ to rate; this could be either an analytical mapping, such as~(\ref{eq:SC_rate}), or a mapping table representing the actual performance of {LTE}. In our simulations, we employ the Bit-Interleaved Coded-Modulation (BICM) capacity as proposed in~\cite{Sch11}, since {LTE} is based on a {BICM} architecture.  
	\item Determine the optimal {LTE} transmission rates per layer for each $L$ and $\hat{\Mat{W}}(L)$. We employ a target Block Error Ratio (BLER) mapping in our simulations to determine the highest rate that achieves $\text{BLER} \leq 0.1$. 
	\item Select the transmission rank $\hat{L}$ that maximizes the sum rate over spatial layers, utilizing the {LTE} transmission rates determined above. 
	\item Set the {RI} and {PMI} according to $\hat{L}$ and $\hat{\Mat{W}}(L)$, respectively and set the {CQI}s conforming to the corresponding {LTE} transmission rates.    
\end{enumerate}     

In Figure~\ref{fig:TP_4x4}, we evaluate the performance of single-user {MIMO} {LTE} uplink transmission over $N_T = N_R = 4$ antennas with link adaptation, $1.4\,$MHz system bandwidth and {ZF} receiver. We do not consider signalling delays between the base station and the user. We employ the VehA channel model~\cite{tr25943} and compare the absolute and relative (to channel capacity) throughput to the performance bounds proposed in~\cite{Schwarz-bounds2011}.\footnote{Notice that the simulation setup is the same as employed in~\cite{Schwarz-bounds2011} for the investigation of {LTE} downlink transmission, thus, facilitating the comparison of up- and downlink performance.} \emph{Channel capacity} is obtained by  applying Singular Value Decomposition (SVD)-based transceivers and water-filling power allocation over subcarriers and spatial streams. Notice that we do not account for guard band and {CP} overheads when calculating the channel capacity; that is, we only consider subcarriers that are available for data transmission. The \emph{achievable channel capacity} takes overhead for pilot symbols ({DMRS} and {SRS}) into account, corresponding to a loss of 16.7\,\% in our simulation. The \emph{achievable {BICM} bound} additionally accounts for equal power allocation, codebook-based precoding, {ZF} detection as well as the applied {BICM} architecture as detailed in~\cite{Schwarz-bounds2011}. 

The performance of {LTE} uplink transmission with full link adaptation (\emph{PMI and rank adaptive}) is similar to the \emph{achievable {BICM} bound} but shifted by approximately 3\,dB. Notice that the saturation value is not the same because the highest {CQI} of {LTE} achieves 5.55\,bit/channel use, whereas the {BICM} bound saturates at 6\,bit/channel use. We also show the performance of {LTE} uplink when restricted to fixed precoding (\emph{rank adaptive}) and fixed rank transmission (\emph{rank 1, 2, 3, 4}). We observe that rank adaptive transmission even outperforms the envelope of the fixed rank transmission curves, since instantaneous rank adaptation selects the optimal rank in each subframe independently. In terms of relative throughput, we see that {LTE} uplink with {ZF} receivers achieves around 40-50\,\% of channel capacity; remember, though, that this does not include {CP} and guard band overheads.

%
%
%
%
%
%


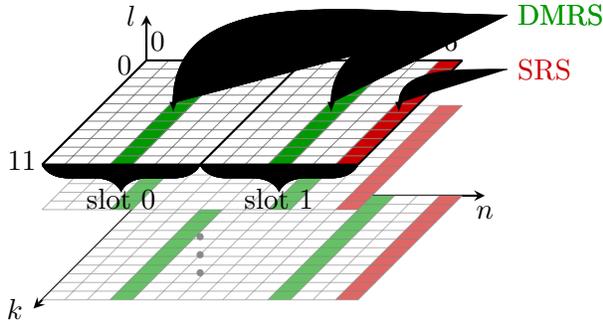
\begin{figure}[t!]
	\begin{center}
\begin{tikzpicture}[>=stealth,scale=0.6,grid/.style={very thin,gray}, 3d/.style={canvas is xy plane at z=0}, border/.style={thick, black}]
\usetikzlibrary{decorations.pathreplacing}

\def\layers{3};

\draw[->,thick] (-0.25,0,0) -- (7.5,0,0) node[anchor=north]{$n$};
\draw[->,thick] (0,-0.25,0) -- (0,4,0) node[anchor=east]{$l$};
\draw[->,thick] (0,0,-0.25) -- (0,0,6.5) node[anchor=east]{$k$};

\foreach \y in {0,2,3}
{
	\ifthenelse{\y<3}
	{\def\shade{0.6};}
	{\def\shade{1};};
	
	\begin{scope}[opacity=\shade]
		\fill[canvas is xz plane at y=\y,white,opacity=1]	     (0,0) 		rectangle (7,6);
		\fill[canvas is xz plane at y=\y,green!60!black] (1.5,0)	rectangle (2,6);
		\fill[canvas is xz plane at y=\y,green!60!black] (5,0) 		rectangle (5.5,6);
		\fill[canvas is xz plane at y=\y,red!80!black] 	 (6.5,0) 	rectangle (7,6);
		\draw[step=0.5cm,grid,canvas is xz plane at y=\y] (0,0) grid (7,6);
		
		\ifthenelse{\y=3}
		{	
			\draw[border] (0,\y,0) 	 -- (0,\y,6);
			\draw[border] (7,\y,0)   -- (7,\y,6);
			\draw[border] (0,\y,0)   -- (7,\y,0);
			\draw[border] (0,\y,6)   -- (7,\y,6);
			\draw[thick] (3.5,\y,0) -- (3.5,\y,6);

	
			\draw (0.25,\y,0) node [anchor=south]{0};
			\draw (3.25,\y,0) node [anchor=south]{6};
			\draw (3.75,\y,0) node [,anchor=south]{0};
			\draw (6.75,\y,0) node [anchor=south]{6};
			\draw (0,\y,5.75) node [anchor=east]{11};
			\draw (0,\y,0.25) node [anchor=east]{0};
			\draw [canvas is xz plane at y=\y,decorate,decoration={brace,mirror,amplitude=10pt},xshift=0pt,yshift=0cm] (0,6) -- (3.5,6) node [black,midway,yshift=-0.45cm] {slot 0};
			\draw [canvas is xz plane at y=\y,decorate,decoration={brace,mirror,amplitude=10pt},xshift=0pt,yshift=0cm] (3.5,6) -- (7,6) node [black,midway,yshift=-0.45cm] {slot 1};
			\draw[-latex] (8,\y+1,0) node [green!60!black,right]{DMRS}to[out=180,in=90] (1.75,\y,3);
			\draw[-latex] (8,\y+1,0) node [green!60!black,right]{DMRS}to[out=180,in=90] (5.25,\y,3);
			\draw[-latex] (8,\y-0.2,0) node [red!80!black,right]{SRS}to[out=180,in=90] (6.75,\y,3);
		}
		
	\end{scope}
}

\fill[grid,opacity=0.9] (3.5,1.4,6) circle [radius=0.08cm];
\fill[grid,opacity=0.9] (3.5,1,6) 	circle [radius=0.08cm];
\fill[grid,opacity=0.9] (3.5,0.6,6) circle [radius=0.08cm];

\end{tikzpicture}
		\caption{The LTE-A uplink reference symbol allocation in two slots (one subframe).}
		\label{fig:CE_resourcegrid}
	\end{center}
\end{figure}

\section{Reference Symbols}
\label{sec:CE_reference_symbols}
In {LTE} uplink two types of reference signals are standardized. For {CE} and coherent detection, {DMRS} are exploited, while {SRS} are employed for channel sounding to enable frequency selective scheduling. For the purpose of {CE} we will consider {DMRS} only. The reference symbols are defined in~\cite{3GPP_TS_36211} and are explained in more detail in~\cite{Hou:dmrsdesign,Zhang:optimizing}. As shown in Figure\ref{fig:CE_resourcegrid}, {DMRS} are multiplexed in the resource grid at {OFDM} symbol time $n=3$ in every slot. In a Physical Uplink Shared Channel (PUSCH) transmission of the {LTE-A} uplink, a {DMRS} occupies all scheduled subcarriers. We assume that the user is assigned all $N_{\rm SC}$ subcarriers starting at $0$, i.e., $k\in\lbrace 0,1,\hdots,N_{\rm SC}-1\rbrace$. We denote the Zadoff-Chu (ZC) base sequence on $N_{\rm SC}$ subcarriers for one slot by $\Vec{\bar{r}} \in \mathbb{C}^{N_{\rm SC} \times 1}$. The base sequences $\Vec{\bar{r}}$ are complex exponential sequences lying on the unit circle fulfilling
\begin{equation} 
\left| \left[ \Vec{\bar{r}} \right]_k \right| = 1~.
\label{eqn:CE_zc}
\end{equation}
In {LTE-A} the {DMRS} of different transmission layers in the same slot are orthogonal in terms of Frequency Domain Code Division Multiplexing (FD-CDM)~\cite{Hou:dmrsdesign}. This is obtained by cyclically shifting the base sequence. Similar to~\cite{Chen:channelestimation}, {DMRS} on layer $l$ for one slot are given by
\begin{equation}
\Mat{R}^{(l)}= \Diag \big( \Vec{r}^{(l)} \big) = \Mat{T}^{(l)} \Diag{\left(\Vec{\bar{r}}\right)}~,
\label{eqn:CE_def_dmrs}
\end{equation}
with the cyclic shift operator
\begin{equation}
\Mat{T}^{(l)}=\Diag\left(e^{j0},\hdots,e^{j\alpha_l k},\hdots,e^{j\alpha_l (N_{\rm SC}-1)}\right)~,
\label{eqn:CE_cyclic_shift}
\end{equation}
and the layer dependent cyclic shift $\alpha_l$. We further conclude from~(\ref{eqn:CE_zc})-(\ref{eqn:CE_cyclic_shift}) that $(\Mat{R}^{(l)})^H = (\Mat{R}^{(l)})^{-1}$ which implies $\big(\Mat{R}^{(l)}\big)^H \Mat{R}^{(l)} = \Mat{I}_{N_{\rm SC}}$. Exploiting~(\ref{eqn:CE_zc}), the product of two {DMRS} from layers $l$ and $u$ with $l,u\in\lbrace 1,\hdots,L\rbrace$, becomes
\begin{eqnarray}
\big(\Mat{R}^{(l)}\big)^H \Mat{R}^{(u)} 	 = \big( \Mat{T}^{(l)} \big)^H \Mat{T}^{(u)}  \Diag{\left(\Vec{\bar{r}}\right)}^H \Diag{\left(\Vec{\bar{r}}\right)}\notag\\
											 = \Diag \left( e^{j0} \;\hdots\; e^{j\Delta\alpha k} \;\hdots\; e^{j\Delta\alpha (N_{\rm SC}-1)} \right) \Mat{I} ~,
\label{eqn:CE_dmrs_times_dmrs}
\end{eqnarray}
with $\Delta \alpha = \alpha_u - \alpha_l$ being the cyclic phase shift between {DMRS} of two different spatial layers. The {FD-CDM} orthogonality can therefore be exploited as
\begin{equation}
\trace \!\! \left( \big(\Mat{R}^{(u)}\big)^H \Mat{R}^{(l)} \right) \! = \! \big(\Vec{r}^{(u)}\big)^H \Vec{r}^{(l)} \! = \begin{cases}	N_{\rm SC} 	&\mbox{for } u=l  \\ 
																														0 			&\mbox{for } u\neq l ~. \end{cases}
\end{equation}
After transmission over a frequency selective channel, this orthogonality has to be exploited to separate all effective {MIMO} channels at the receiver.

\section{Channel Estimation}
\label{sec:CE_channel_estimation}
For channel estimation we exploit the system model only at symbol times, where reference signals are allocated. For normal {CP} length this is the 4\textsuperscript{th} symbol in each slot, i.e., $n=3$ as shown in Figure \ref{fig:CE_resourcegrid}. Since we estimate the channel only at this single symbol time per slot, interpolation in time has to be carried out to obtain channel estimates for the whole resource grid. The effects of interpolation will be studied in Section~\ref{sec:channel_interpolation}. As illustrated in Figure~\ref{fig:system_model}, the {DMRS} are added after {DFT} spreading, right before precoding. As the channel estimation takes place after the receiver's {DFT}, just before equalization, the system model for {CE} amounts to an {OFDM} system. The system model~(\ref{eq:IO}) therefore reads as
\begin{equation}
\label{eqn:CE_system_model_2}
\Vec{y} 			= \Mat{H}_{\rm eff} \Vec{r} + \Vec{n'} ~,
\end{equation}
with (pre-equalization) noise
\begin{equation}
\label{eqn:CE_pre_equal_noise}
\Vec{n'} = \left( \Mat{I}_{N_{R}} \otimes \Mat{M}^H \Mat{D}_{N_{FFT}} \Mat{P}_{\text{remCP}}\right) \Vec{n} ~,
\end{equation}
and the stacked vector $\Vec{r}$ consisting of {DMRS} $\Vec{r}^{(l)} \in \mathbb{C}^{N_{\rm SC} \times 1}$ from all active spatial layers $l\in \lbrace 1,\dots,L \rbrace$, i.e. $\Vec{r} = \big( \big(\Vec{r}^{(1)}\big)^T,\dots,\big(\Vec{r}^{(L)}\big)^T\big)^T$. To consider the received signal separately for each receive antenna $i$, we can select the according part from $\Vec{y}$ by left multiplying with the selector matrix $\Mat{S}^{(i)}$ from~(\ref{eq:selector_matrix}). The received signal $\Vec{y}^{(i)} = \Mat{S}^{(i)} \Vec{y}$ on antenna $i$ is given by
\begin{align}
\label{eqn:CE_system_model_2}
\Vec{y}^{(i)} 	= & \left( \Mat{H}_{\rm eff}^{(i,1)} ,\dots, \Mat{H}_{\rm eff}^{(i,L)} \right) \Vec{r} + \Vec{n'}^{(i)} \\
				= & \sum_{l=1}^L 	\Mat{H}_{\rm eff}^{(i,l)} \Vec{r}^{(l)} + \Vec{n'}^{(i)} ~, \notag
\end{align}
with the pre-equalization noise $\Vec{n'}^{(i)} = \Mat{S}^{(i)} \Vec{n'}$ on receive antenna $i$ and $\Mat{H}^{(i,l)}_{\rm eff} = \Mat{S}^{(i)} \Mat{H}_{\rm eff} \big(\Mat{S}^{(l)}\big)^T$ being the $(i,l)^{\rm th}$ block of $\Mat{H}_{\rm eff}$. Since $\Mat{H}^{(i,l)}_{\rm eff}$ is diagonal, we exploit the relations  $\Mat{R}^{(l)} = \Diag \big( \Vec{r}^{(l)} \big)$ and $\Vec{h}_{\rm eff}^{(i,l)} = \diag \big( \Mat{H}_{\rm eff}^{(i,l)} \big)$ to estimate a channel vector rather than a matrix and rearrange terms in~(\ref{eqn:CE_system_model_2}) leading to
\begin{align}
\label{eqn:CE_system_model_3}
\Vec{y}^{(i)}	= & \sum_{l=1}^L \Mat{R}^{(l)} \Vec{h}_{\rm eff}^{(i,l)} + \Vec{n'}^{(i)} \\
				= & \underbrace{\left( \Mat{R}^{(1)} ,\dots, \Mat{R}^{(L)} \right)}_{\Mat{R}} \Vec{h}_{\rm eff}^{(i)} + \Vec{n'}^{(i)} ~, \notag
\end{align}
with the stacked vector $\Vec{h}_{\rm eff}^{(i)} = \big( \big(\Vec{h}_{\rm eff}^{(i,1)}\big)^T,\dots,\big(\Vec{h}_{\rm eff}^{(i,L)}\big)^T\big)^T$ of all effective channels from $L$ active layers to receive antenna $i$ for which we will drop the subscript in the following.\\

\subsection{Minimum Mean Square Error Estimation}
\label{sec:sub:CE_MMSE}
First we present a {MMSE} estimator where we exploit~(\ref{eqn:CE_system_model_3}) and estimate the stacked vector $\Vec{h}^{(i)}$ consisting of effective channels from all $L$ active layers to receive antenna $i$. The {MMSE} {CE} for receive antenna $i$ is given by
\begin{equation}
\Vec{\hat{h}}^{(i)}_{\rm MMSE} = \argmin\limits_{\Vec{\hat{h}}^{(i)}} \mathbb{E} \left\lbrace \big\|  \Vec{\hat{h}}^{(i)} - \Vec{h}^{(i)} \big\|_2^2 \right\rbrace ~,
\label{eqn:CE_MMSE}
\end{equation}
which leads to the well-known solution~\cite{kay1993fundamentals}
\begin{equation}
\label{eqn:CE_MMSE_estimator}
\Vec{\hat{h}}^{(i)}_{\rm MMSE} = \left( \sigma_{\Vec{n}^{(i)}}^2 \Mat{C}_{\Vec{h}^{(i)}}^{-1} + \Mat{R}^{H} \Mat{R}\right)^{-1} \Mat{R}^{H} \Vec{y}^{(i)} ~,
\end{equation}
with $\Mat{C}_{\Vec{h}^{(i)}} = \mathbb{E} \lbrace \Vec{h}^{(i)} \Vec{h}^{(i)H}\rbrace$.\\

\begin{figure*}
\centering
	\subfigure[Mean Squared Error (MSE) curves of the proposed {CE} methods for a $2\times 2$ transmission with $L=2$ on a Typical Urban (TU) channel.]{\includegraphics[width=0.45\textwidth]{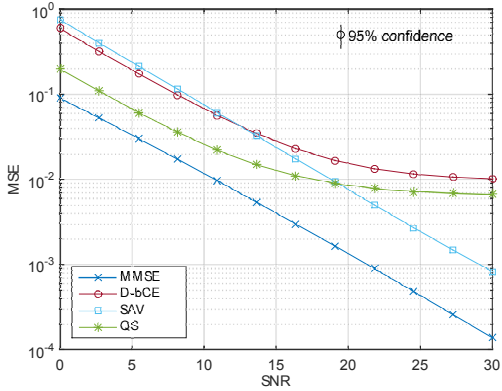} \label{fig:CE_performance_MSE}}\qquad
	\subfigure[Resulting uncoded {BER} curves for the proposed {CE} schemes for 4QAM.]{\includegraphics[width=0.45\textwidth]{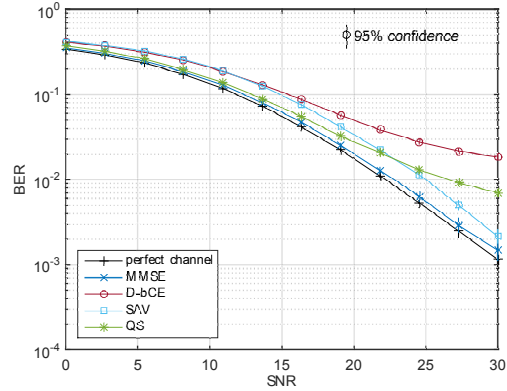} \label{fig:CE_performance_BER}}
	\caption{Channel estimation performance comparison for block fading.}
	\label{fig:CE_performance}
\end{figure*}

\subsection{Correlation Based Estimation}
\label{sec:sub:CE_correlation_based_estimation}
As a low complexity approach, we correlate (matched filter) the received signal with the reference symbol of layer $l$ to obtain a channel estimate for the effective channel $\Vec{h}^{(i,l)}$ from layer $l$ to receive antenna $i$
\begin{equation}
\Vec{\tilde{h}}^{(i,l)} 	= \big(\Mat{R}^{(l)}\big)^H \Vec{y}^{(i)} ~.
\label{eqn:CE_correlation}
\end{equation}
Inserting our system model~(\ref{eqn:CE_system_model_3}) and exploiting~(\ref{eqn:CE_dmrs_times_dmrs}), we obtain
\begin{eqnarray}
\label{eqn:CE_correlation}
\hspace{-6ex} \Vec{\tilde{h}}^{(i,l)}	= & \big(\Mat{R}^{(l)}\big)^H \sum_{u=1}^{L}  \Mat{R}^{(u)} \Vec{h}^{(i,u)} + \big(\Mat{R}^{(l)}\big)^H \Vec{n'}^{(i)} \\
	\hspace{-6ex}						= & \Vec{h}^{(i,l)} + \underbrace{\sum_{\substack{u=1\\ u\neq l}}^{L} \big(\Mat{T}^{(l)}\big)^H \Mat{T}^{(u)} \Vec{h}^{(i,u)}}_{\text{inter-layer interference}} + \; \Vec{\tilde{n}}^{(i)} \notag ~.
\end{eqnarray}
Here $\Vec{\tilde{n}}^{(i)}$ has the same distribution as $\Vec{n'}^{(i)}$ since $(\Mat{R}^{(l)})^H$ is unitary and introduces phase changes only, cf.~(\ref{eqn:CE_def_dmrs}). Due to the allocation of {DMRS} on the same time and frequency resources on different spatial layers, the initial estimate $\Vec{\tilde{h}}^{(i,l)}$ of one effective {MIMO} channel actually consists of a superposition of all $L$ effective MIMO channels to receive antenna $i$. The unintentional contributions in~(\ref{eqn:CE_correlation}), from layers $u\neq l$ are inter-layer interference, making it unsuited as initial estimate for coherent detection. Different methods to separate the different effective {MIMO} channels in~(\ref{eqn:CE_correlation}) will be presented in the following.\\

\subsubsection{DFT based Channel Estimation}
\label{sec:sub:CE_dft_based_channel_stimation}
A well known approach for {CE} in {LTE-A} uplink is {DFT} based estimation~\cite{Zhang:optimizing}, which aims to separate the {MIMO} channels contributing to~(\ref{eqn:CE_correlation}) in time domain. For this the individual cyclic shift of each {DMRS} is exploited. Applying a {DFT} on the receive signal, the individual phase shifts will translate into shifts in time domain. This makes a separation of Channel Impulse Responses (CIR)s from different {MIMO} channels possible by windowing. In our simulator we implemented a {DFT} based estimator as in~\cite{zhang:enhanceddft} or~\cite{Chen:channelestimation}.\\

\subsubsection{Averaging}
\label{sec:sub:CE_averaging}
For physically meaningful channels, neighbouring subcarriers will be correlated within the coherence bandwidth~\cite{sklar1997rayleigh}. We utilize this property and exploit the {DMRS} structure to perform frequency domain {CE}. As explained in \cite{pratschner2015uplinkCE}, applying a sliding averaging on the initial estimate $\Vec{\tilde{h}}^{(i,l)}$ from~(\ref{eqn:CE_correlation}) over $\bar{\gamma}$ adjacent subcarriers ($\bar{\gamma}$ equals $1,2,4,4$ for $L$ equals $1,2,3,4$, respectively) cancels the inter-layer interference, assuming the channel to be frequency flat on these $\bar{\gamma}$ consecutive subcarriers. The sliding average is given by
\begin{equation}
\left[ \Vec{\hat{h}}_{\rm SAV}^{(i,l)} \right]_k = \frac{1}{\bar{\gamma} ^2} \sum_{t= k-\bar{\gamma} +1}^{k} \sum_{j=t}^{t+\bar{\gamma}-1} \left[ \Vec{\tilde{h}}^{(i,l)} \right]_j ~,
\label{eqn:CE_SAV_estimator}
\end{equation}
for $\bar{\gamma} \leq k \leq N_{\rm SC}-\bar{\gamma} +1$. The second sum describes the averaging of $\bar{\gamma}$ elements while the first sum describes the shift of this averaging window.\\

\subsubsection{Quadratic Smoothing}
\label{sec:sub:CE_quadratic_smoothing}
Another method exploiting channel correlations to estimate the channel in frequency domain is Quadratic Smoothing (QS). This scheme cannot remove the inter-layer interference entirely, which manifests in a higher error floor, but shows improved performance at lower {SNR} in return. As explained in~\cite{pratschner2015uplinkCE} this estimation method, exploiting the smoothing matrix $\Mat{Q}$ and a smoothing factor $\gamma$, is given by
\begin{align}
\label{eqn:CE_QS_estimator}
\Vec{\hat{h}}_{\rm QS}^{(i,l)} = \left( \Mat{I}_{N_{\rm SC}L} + \lambda  \Mat{Q}^H\Mat{Q}  \right)^{-1} \underbrace{\big(\Mat{{R}}^{(l)}\big)^{H} \Vec{{y}}^{(i)}}_{\Vec{\tilde{h}}^{(i,l)}} ~.
\end{align}
Similar to~(\ref{eqn:CE_SAV_estimator}) this can be interpreted as another way to cope with the inter-layer interference in~(\ref{eqn:CE_correlation}) by post processing. This method does not use the {DMRS} structure explicitly but suppresses the interference by smoothing. It is therefore not able to cancel the complete inter-layer interference but shows a improved performance at low SNR.\\

\subsection{{MSE} and {BER} comparison}
\label{sec:CE_simulations}
We assume a single user $2\times 2$ {MIMO} transmission with $N_{\rm SC}=72$ subarriers, a fixed number of layers $L=2$ and a {TU} channel model~\cite{tr25943} at zero speed. We perform a simulation with one point extrapolation, cf.Section~\ref{sec:channel_interpolation}, and show the {MSE} curves of the proposed estimators in Figure~\ref{fig:CE_performance_MSE}. The {DFT} based {CE} (\textit{D-bCE}) shows the highest error flow of all estimators at high {SNR} while the \textit{MMSE} estimator, of course shows best performance over the whole {SNR} range. Compared to these two methods, the Sliding-Averaging estimator~(\ref{eqn:CE_SAV_estimator}), denoted by \textit{SAV}, encounters a 8dB {SNR} penalty when compared to \textit{MMSE}, but comes closest to \textit{MMSE} performance at high {SNR}. The quadratic smoothing estimation is denoted by \textit{QS} and shows a significant improvement for low {SNR}s because it smooths over several observed channel coefficients. Quadratic smoothing performs uniformly better than \textit{D-bC}E over the whole {SNR} range and comes close to 4dB to \textit{MMSE} at low {SNR}. The high error floor shows that \textit{QS} is not able to cancel all the inter-layer interference.\\
In terms of {BER} performance, at high {SNR}, naturally the estimation method with lowest {MSE} leads to the smallest {BER}. At low {SNR}, the difference in {CE} {MSE} translates into very small differences in {BER}, meaning we cannot gain too much from a good low {SNR} MSE performance of \textit{QS} or \textit{MMSE} estimation. Considering estimation complexity and that \textit{MMSE} as well as \textit{QS} require prior channel knowledge, \textit{SAV} estimation is a good complexity performance trade-off.\\


\section{Channel Interpolation}
\label{sec:channel_interpolation}

\begin{figure}[h!]
  \centering
  \subfigure[Resource grid]{
  \begin{tikzpicture}[>=latex,scale=0.85,font=\footnotesize] 
        \newcommand{\boxlen}{0.2};
  \newcommand{\pilotCol}{red}
  \newcommand{\pilotColLight}{red!40!white}
  \newcommand{\lineColLight}{black!40!white}

  \draw[draw=none,fill=\pilotColLight] (\boxlen*4-\boxlen,0) rectangle (\boxlen*4,\boxlen*12);
  \draw[draw=none,fill=\pilotColLight] (\boxlen*11-\boxlen,0) rectangle (\boxlen*11,\boxlen*12);
  \draw[draw=none,fill=\pilotCol] (\boxlen*18-\boxlen,0) rectangle (\boxlen*18,\boxlen*12);
  \draw[draw=none,fill=\pilotCol] (\boxlen*25-\boxlen,0) rectangle (\boxlen*25,\boxlen*12);

 \foreach \x in {0,...,13} {
    \draw[very thin,\lineColLight] (\x*\boxlen,0) -- (\x*\boxlen,12*\boxlen);
  }
  \foreach \x in {14,...,28} {
    \draw[very thin] (\x*\boxlen,0) -- (\x*\boxlen,12*\boxlen);
  }  
  
  \foreach \y in {0,...,12} {
    \draw[very thin,\lineColLight] (0,\y*\boxlen) -- (14*\boxlen,\y*\boxlen);    
    \draw[very thin] (14*\boxlen,\y*\boxlen) -- (28*\boxlen,\y*\boxlen);
  }

  \draw[ultra thin,\lineColLight] (0*\boxlen,12*\boxlen) -- (0*\boxlen,12*\boxlen+1.7*\boxlen);
  \draw[ultra thin] (14*\boxlen,12*\boxlen) -- (14*\boxlen,12*\boxlen+1.7*\boxlen);
  \draw[ultra thin] (28*\boxlen,12*\boxlen) -- (28*\boxlen,12*\boxlen+1.7*\boxlen);
  \draw[ultra thin,<->,\lineColLight] (0*\boxlen,12*\boxlen+1.3*\boxlen) -- node[above] {subframe u-1} (14*\boxlen,12*\boxlen+1.3*\boxlen);
  \draw[ultra thin,<->] (14*\boxlen,12*\boxlen+1.3*\boxlen) -- node[above] {subframe u} (28*\boxlen,12*\boxlen+1.3*\boxlen);

  \draw[ultra thin] (14*\boxlen,0*\boxlen) -- (14*\boxlen,-1.7*\boxlen);
  \draw[ultra thin] (21*\boxlen,0*\boxlen) -- (21*\boxlen,-1.7*\boxlen);
  \draw[ultra thin] (28*\boxlen,0*\boxlen) -- (28*\boxlen,-1.7*\boxlen);

  \draw[ultra thin,<->] (14*\boxlen,-1.3*\boxlen) -- node[below] {1. slot} (21*\boxlen,-1.3*\boxlen);
  \draw[ultra thin,<->] (21*\boxlen,-1.3*\boxlen) -- node[below] {2. slot} (28*\boxlen,-1.3*\boxlen);

   \draw[very thin,->]  (0.5*\boxlen,-1*\boxlen) -- node[below] {time} (4.5*\boxlen,-1*\boxlen);
   \draw[very thin,->]  (-\boxlen,3*\boxlen) -- node[above,rotate=90] {frequency} (-\boxlen,9*\boxlen);

  \begin{scope}[xshift=\boxlen*30.0 cm, yshift=3*\boxlen cm]
     \draw[draw=black,very thin] (0,\boxlen*11-\boxlen) rectangle (\boxlen,\boxlen*11);
     \draw[draw=black,very thin,fill=\pilotCol] (0,\boxlen*9-\boxlen) rectangle (\boxlen,\boxlen*9);
     \node[right] at (\boxlen,\boxlen*11-0.5*\boxlen) {data symbol};
     \node[right] at (\boxlen,\boxlen*9-0.5*\boxlen) {pilot symbol};
  \end{scope} 
  \end{tikzpicture}   
  }\\

  \subfigure[1 point extrapolation]{
  \begin{tikzpicture}[>=latex,scale=1,font=\footnotesize] 
      \draw[-,black!40!white] (-1.716000,0) -- (-1.716000,1.183680);
\draw[fill=black!40!white,draw=none] (-1.716000,1.183680) circle (0.05);
\draw[-,black!40!white] (-1.584000,0) -- (-1.584000,1.395720);
\draw[fill=black!40!white,draw=none] (-1.584000,1.395720) circle (0.05);
\draw[-,black!40!white] (-1.452000,0) -- (-1.452000,1.291080);
\draw[fill=black!40!white,draw=none] (-1.452000,1.291080) circle (0.05);
\draw[-,red] (-1.320000,0) -- (-1.320000,1.196160);
\draw[fill=red,draw=none] (-1.320000,1.196160) circle (0.05);
\draw[-,black!40!white] (-1.188000,0) -- (-1.188000,1.442280);
\draw[fill=black!40!white,draw=none] (-1.188000,1.442280) circle (0.05);
\draw[-,black!40!white] (-1.056000,0) -- (-1.056000,1.111080);
\draw[fill=black!40!white,draw=none] (-1.056000,1.111080) circle (0.05);
\draw[-,black!40!white] (-0.924000,0) -- (-0.924000,1.155960);
\draw[fill=black!40!white,draw=none] (-0.924000,1.155960) circle (0.05);
\draw[-,black!40!white] (-0.792000,0) -- (-0.792000,1.119480);
\draw[fill=black!40!white,draw=none] (-0.792000,1.119480) circle (0.05);
\draw[-,black!40!white] (-0.660000,0) -- (-0.660000,1.233960);
\draw[fill=black!40!white,draw=none] (-0.660000,1.233960) circle (0.05);
\draw[-,black!40!white] (-0.528000,0) -- (-0.528000,1.273320);
\draw[fill=black!40!white,draw=none] (-0.528000,1.273320) circle (0.05);
\draw[-,red] (-0.396000,0) -- (-0.396000,1.114920);
\draw[fill=red,draw=none] (-0.396000,1.114920) circle (0.05);
\draw[-,black!40!white] (-0.264000,0) -- (-0.264000,1.119120);
\draw[fill=black!40!white,draw=none] (-0.264000,1.119120) circle (0.05);
\draw[-,black!40!white] (-0.132000,0) -- (-0.132000,1.212840);
\draw[fill=black!40!white,draw=none] (-0.132000,1.212840) circle (0.05);
\draw[-,black!40!white] (0.000000,0) -- (0.000000,1.029600);
\draw[fill=black!40!white,draw=none] (0.000000,1.029600) circle (0.05);
\draw[-,black] (0.132000,0) -- (0.132000,1.238520);
\draw[fill=black,draw=none] (0.132000,1.238520) circle (0.05);
\draw[-,black] (0.264000,0) -- (0.264000,1.223280);
\draw[fill=black,draw=none] (0.264000,1.223280) circle (0.05);
\draw[-,black] (0.396000,0) -- (0.396000,1.107240);
\draw[fill=black,draw=none] (0.396000,1.107240) circle (0.05);
\draw[-,red] (0.528000,0) -- (0.528000,1.277160);
\draw[fill=red,draw=none] (0.528000,1.277160) circle (0.05);
\draw[-,black] (0.660000,0) -- (0.660000,1.045680);
\draw[fill=black,draw=none] (0.660000,1.045680) circle (0.05);
\draw[-,black] (0.792000,0) -- (0.792000,1.070640);
\draw[fill=black,draw=none] (0.792000,1.070640) circle (0.05);
\draw[-,black] (0.924000,0) -- (0.924000,1.124520);
\draw[fill=black,draw=none] (0.924000,1.124520) circle (0.05);
\draw[-,black] (1.056000,0) -- (1.056000,1.210560);
\draw[fill=black,draw=none] (1.056000,1.210560) circle (0.05);
\draw[-,black] (1.188000,0) -- (1.188000,0.944160);
\draw[fill=black,draw=none] (1.188000,0.944160) circle (0.05);
\draw[-,black] (1.320000,0) -- (1.320000,1.303320);
\draw[fill=black,draw=none] (1.320000,1.303320) circle (0.05);
\draw[-,red] (1.452000,0) -- (1.452000,1.071120);
\draw[fill=red,draw=none] (1.452000,1.071120) circle (0.05);
\draw[-,black] (1.584000,0) -- (1.584000,1.332720);
\draw[fill=black,draw=none] (1.584000,1.332720) circle (0.05);
\draw[-,black] (1.716000,0) -- (1.716000,1.273320);
\draw[fill=black,draw=none] (1.716000,1.273320) circle (0.05);
\draw[-,black] (1.848000,0) -- (1.848000,1.219560);
\draw[fill=black,draw=none] (1.848000,1.219560) circle (0.05);
\draw[->] (-1.848000,0) -- (2.112000,0) node[at end, below] {};
\draw[thin,red] (0.066000,1.277160) -- (0.990000,1.277160);
\draw[thin,red] (0.990000,1.071120) -- (1.914000,1.071120); 
  \end{tikzpicture}   
 }
\subfigure[2 point linear interpolation]{
  \begin{tikzpicture}[>=latex,scale=1,font=\footnotesize] 
      \draw[-,black!40!white] (-1.716000,0) -- (-1.716000,1.183680);
\draw[fill=black!40!white,draw=none] (-1.716000,1.183680) circle (0.05);
\draw[-,black!40!white] (-1.584000,0) -- (-1.584000,1.395720);
\draw[fill=black!40!white,draw=none] (-1.584000,1.395720) circle (0.05);
\draw[-,black!40!white] (-1.452000,0) -- (-1.452000,1.291080);
\draw[fill=black!40!white,draw=none] (-1.452000,1.291080) circle (0.05);
\draw[-,red] (-1.320000,0) -- (-1.320000,1.196160);
\draw[fill=red,draw=none] (-1.320000,1.196160) circle (0.05);
\draw[-,black!40!white] (-1.188000,0) -- (-1.188000,1.442280);
\draw[fill=black!40!white,draw=none] (-1.188000,1.442280) circle (0.05);
\draw[-,black!40!white] (-1.056000,0) -- (-1.056000,1.111080);
\draw[fill=black!40!white,draw=none] (-1.056000,1.111080) circle (0.05);
\draw[-,black!40!white] (-0.924000,0) -- (-0.924000,1.155960);
\draw[fill=black!40!white,draw=none] (-0.924000,1.155960) circle (0.05);
\draw[-,black!40!white] (-0.792000,0) -- (-0.792000,1.119480);
\draw[fill=black!40!white,draw=none] (-0.792000,1.119480) circle (0.05);
\draw[-,black!40!white] (-0.660000,0) -- (-0.660000,1.233960);
\draw[fill=black!40!white,draw=none] (-0.660000,1.233960) circle (0.05);
\draw[-,black!40!white] (-0.528000,0) -- (-0.528000,1.273320);
\draw[fill=black!40!white,draw=none] (-0.528000,1.273320) circle (0.05);
\draw[-,red] (-0.396000,0) -- (-0.396000,1.114920);
\draw[fill=red,draw=none] (-0.396000,1.114920) circle (0.05);
\draw[-,black!40!white] (-0.264000,0) -- (-0.264000,1.119120);
\draw[fill=black!40!white,draw=none] (-0.264000,1.119120) circle (0.05);
\draw[-,black!40!white] (-0.132000,0) -- (-0.132000,1.212840);
\draw[fill=black!40!white,draw=none] (-0.132000,1.212840) circle (0.05);
\draw[-,black!40!white] (0.000000,0) -- (0.000000,1.029600);
\draw[fill=black!40!white,draw=none] (0.000000,1.029600) circle (0.05);
\draw[-,black] (0.132000,0) -- (0.132000,1.238520);
\draw[fill=black,draw=none] (0.132000,1.238520) circle (0.05);
\draw[-,black] (0.264000,0) -- (0.264000,1.223280);
\draw[fill=black,draw=none] (0.264000,1.223280) circle (0.05);
\draw[-,black] (0.396000,0) -- (0.396000,1.107240);
\draw[fill=black,draw=none] (0.396000,1.107240) circle (0.05);
\draw[-,red] (0.528000,0) -- (0.528000,1.277160);
\draw[fill=red,draw=none] (0.528000,1.277160) circle (0.05);
\draw[-,black] (0.660000,0) -- (0.660000,1.045680);
\draw[fill=black,draw=none] (0.660000,1.045680) circle (0.05);
\draw[-,black] (0.792000,0) -- (0.792000,1.070640);
\draw[fill=black,draw=none] (0.792000,1.070640) circle (0.05);
\draw[-,black] (0.924000,0) -- (0.924000,1.124520);
\draw[fill=black,draw=none] (0.924000,1.124520) circle (0.05);
\draw[-,black] (1.056000,0) -- (1.056000,1.210560);
\draw[fill=black,draw=none] (1.056000,1.210560) circle (0.05);
\draw[-,black] (1.188000,0) -- (1.188000,0.944160);
\draw[fill=black,draw=none] (1.188000,0.944160) circle (0.05);
\draw[-,black] (1.320000,0) -- (1.320000,1.303320);
\draw[fill=black,draw=none] (1.320000,1.303320) circle (0.05);
\draw[-,red] (1.452000,0) -- (1.452000,1.071120);
\draw[fill=red,draw=none] (1.452000,1.071120) circle (0.05);
\draw[-,black] (1.584000,0) -- (1.584000,1.332720);
\draw[fill=black,draw=none] (1.584000,1.332720) circle (0.05);
\draw[-,black] (1.716000,0) -- (1.716000,1.273320);
\draw[fill=black,draw=none] (1.716000,1.273320) circle (0.05);
\draw[-,black] (1.848000,0) -- (1.848000,1.219560);
\draw[fill=black,draw=none] (1.848000,1.219560) circle (0.05);
\draw[->] (-1.848000,0) -- (2.112000,0) node[at end, below] {};
\draw[thin,red] (0.132000,1.365463) -- (1.848000,0.982817); 
  \end{tikzpicture}   
 }\\
\subfigure[3 point linear interpolation]{
  \begin{tikzpicture}[>=latex,scale=1,font=\footnotesize] 
      \draw[-,black!40!white] (-1.716000,0) -- (-1.716000,1.183680);
\draw[fill=black!40!white,draw=none] (-1.716000,1.183680) circle (0.05);
\draw[-,black!40!white] (-1.584000,0) -- (-1.584000,1.395720);
\draw[fill=black!40!white,draw=none] (-1.584000,1.395720) circle (0.05);
\draw[-,black!40!white] (-1.452000,0) -- (-1.452000,1.291080);
\draw[fill=black!40!white,draw=none] (-1.452000,1.291080) circle (0.05);
\draw[-,red] (-1.320000,0) -- (-1.320000,1.196160);
\draw[fill=red,draw=none] (-1.320000,1.196160) circle (0.05);
\draw[-,black!40!white] (-1.188000,0) -- (-1.188000,1.442280);
\draw[fill=black!40!white,draw=none] (-1.188000,1.442280) circle (0.05);
\draw[-,black!40!white] (-1.056000,0) -- (-1.056000,1.111080);
\draw[fill=black!40!white,draw=none] (-1.056000,1.111080) circle (0.05);
\draw[-,black!40!white] (-0.924000,0) -- (-0.924000,1.155960);
\draw[fill=black!40!white,draw=none] (-0.924000,1.155960) circle (0.05);
\draw[-,black!40!white] (-0.792000,0) -- (-0.792000,1.119480);
\draw[fill=black!40!white,draw=none] (-0.792000,1.119480) circle (0.05);
\draw[-,black!40!white] (-0.660000,0) -- (-0.660000,1.233960);
\draw[fill=black!40!white,draw=none] (-0.660000,1.233960) circle (0.05);
\draw[-,black!40!white] (-0.528000,0) -- (-0.528000,1.273320);
\draw[fill=black!40!white,draw=none] (-0.528000,1.273320) circle (0.05);
\draw[-,red] (-0.396000,0) -- (-0.396000,1.114920);
\draw[fill=red,draw=none] (-0.396000,1.114920) circle (0.05);
\draw[-,black!40!white] (-0.264000,0) -- (-0.264000,1.119120);
\draw[fill=black!40!white,draw=none] (-0.264000,1.119120) circle (0.05);
\draw[-,black!40!white] (-0.132000,0) -- (-0.132000,1.212840);
\draw[fill=black!40!white,draw=none] (-0.132000,1.212840) circle (0.05);
\draw[-,black!40!white] (0.000000,0) -- (0.000000,1.029600);
\draw[fill=black!40!white,draw=none] (0.000000,1.029600) circle (0.05);
\draw[-,black] (0.132000,0) -- (0.132000,1.238520);
\draw[fill=black,draw=none] (0.132000,1.238520) circle (0.05);
\draw[-,black] (0.264000,0) -- (0.264000,1.223280);
\draw[fill=black,draw=none] (0.264000,1.223280) circle (0.05);
\draw[-,black] (0.396000,0) -- (0.396000,1.107240);
\draw[fill=black,draw=none] (0.396000,1.107240) circle (0.05);
\draw[-,red] (0.528000,0) -- (0.528000,1.277160);
\draw[fill=red,draw=none] (0.528000,1.277160) circle (0.05);
\draw[-,black] (0.660000,0) -- (0.660000,1.045680);
\draw[fill=black,draw=none] (0.660000,1.045680) circle (0.05);
\draw[-,black] (0.792000,0) -- (0.792000,1.070640);
\draw[fill=black,draw=none] (0.792000,1.070640) circle (0.05);
\draw[-,black] (0.924000,0) -- (0.924000,1.124520);
\draw[fill=black,draw=none] (0.924000,1.124520) circle (0.05);
\draw[-,black] (1.056000,0) -- (1.056000,1.210560);
\draw[fill=black,draw=none] (1.056000,1.210560) circle (0.05);
\draw[-,black] (1.188000,0) -- (1.188000,0.944160);
\draw[fill=black,draw=none] (1.188000,0.944160) circle (0.05);
\draw[-,black] (1.320000,0) -- (1.320000,1.303320);
\draw[fill=black,draw=none] (1.320000,1.303320) circle (0.05);
\draw[-,red] (1.452000,0) -- (1.452000,1.071120);
\draw[fill=red,draw=none] (1.452000,1.071120) circle (0.05);
\draw[-,black] (1.584000,0) -- (1.584000,1.332720);
\draw[fill=black,draw=none] (1.584000,1.332720) circle (0.05);
\draw[-,black] (1.716000,0) -- (1.716000,1.273320);
\draw[fill=black,draw=none] (1.716000,1.273320) circle (0.05);
\draw[-,black] (1.848000,0) -- (1.848000,1.219560);
\draw[fill=black,draw=none] (1.848000,1.219560) circle (0.05);
\draw[->] (-1.848000,0) -- (2.112000,0) node[at end, below] {};
\draw[thin,red] (-0.396000,1.114920) -- (0.528000,1.277160);
\draw[thin,red] (0.528000,1.277160) -- (1.848000,0.982817); 
  \end{tikzpicture}   
 }
\subfigure[3 point spline interpolation]{
  \begin{tikzpicture}[>=latex,scale=1,font=\footnotesize] 
      \draw[-,black!40!white] (-1.716000,0) -- (-1.716000,1.183680);
\draw[fill=black!40!white,draw=none] (-1.716000,1.183680) circle (0.05);
\draw[-,black!40!white] (-1.584000,0) -- (-1.584000,1.395720);
\draw[fill=black!40!white,draw=none] (-1.584000,1.395720) circle (0.05);
\draw[-,black!40!white] (-1.452000,0) -- (-1.452000,1.291080);
\draw[fill=black!40!white,draw=none] (-1.452000,1.291080) circle (0.05);
\draw[-,red] (-1.320000,0) -- (-1.320000,1.196160);
\draw[fill=red,draw=none] (-1.320000,1.196160) circle (0.05);
\draw[-,black!40!white] (-1.188000,0) -- (-1.188000,1.442280);
\draw[fill=black!40!white,draw=none] (-1.188000,1.442280) circle (0.05);
\draw[-,black!40!white] (-1.056000,0) -- (-1.056000,1.111080);
\draw[fill=black!40!white,draw=none] (-1.056000,1.111080) circle (0.05);
\draw[-,black!40!white] (-0.924000,0) -- (-0.924000,1.155960);
\draw[fill=black!40!white,draw=none] (-0.924000,1.155960) circle (0.05);
\draw[-,black!40!white] (-0.792000,0) -- (-0.792000,1.119480);
\draw[fill=black!40!white,draw=none] (-0.792000,1.119480) circle (0.05);
\draw[-,black!40!white] (-0.660000,0) -- (-0.660000,1.233960);
\draw[fill=black!40!white,draw=none] (-0.660000,1.233960) circle (0.05);
\draw[-,black!40!white] (-0.528000,0) -- (-0.528000,1.273320);
\draw[fill=black!40!white,draw=none] (-0.528000,1.273320) circle (0.05);
\draw[-,red] (-0.396000,0) -- (-0.396000,1.114920);
\draw[fill=red,draw=none] (-0.396000,1.114920) circle (0.05);
\draw[-,black!40!white] (-0.264000,0) -- (-0.264000,1.119120);
\draw[fill=black!40!white,draw=none] (-0.264000,1.119120) circle (0.05);
\draw[-,black!40!white] (-0.132000,0) -- (-0.132000,1.212840);
\draw[fill=black!40!white,draw=none] (-0.132000,1.212840) circle (0.05);
\draw[-,black!40!white] (0.000000,0) -- (0.000000,1.029600);
\draw[fill=black!40!white,draw=none] (0.000000,1.029600) circle (0.05);
\draw[-,black] (0.132000,0) -- (0.132000,1.238520);
\draw[fill=black,draw=none] (0.132000,1.238520) circle (0.05);
\draw[-,black] (0.264000,0) -- (0.264000,1.223280);
\draw[fill=black,draw=none] (0.264000,1.223280) circle (0.05);
\draw[-,black] (0.396000,0) -- (0.396000,1.107240);
\draw[fill=black,draw=none] (0.396000,1.107240) circle (0.05);
\draw[-,red] (0.528000,0) -- (0.528000,1.277160);
\draw[fill=red,draw=none] (0.528000,1.277160) circle (0.05);
\draw[-,black] (0.660000,0) -- (0.660000,1.045680);
\draw[fill=black,draw=none] (0.660000,1.045680) circle (0.05);
\draw[-,black] (0.792000,0) -- (0.792000,1.070640);
\draw[fill=black,draw=none] (0.792000,1.070640) circle (0.05);
\draw[-,black] (0.924000,0) -- (0.924000,1.124520);
\draw[fill=black,draw=none] (0.924000,1.124520) circle (0.05);
\draw[-,black] (1.056000,0) -- (1.056000,1.210560);
\draw[fill=black,draw=none] (1.056000,1.210560) circle (0.05);
\draw[-,black] (1.188000,0) -- (1.188000,0.944160);
\draw[fill=black,draw=none] (1.188000,0.944160) circle (0.05);
\draw[-,black] (1.320000,0) -- (1.320000,1.303320);
\draw[fill=black,draw=none] (1.320000,1.303320) circle (0.05);
\draw[-,red] (1.452000,0) -- (1.452000,1.071120);
\draw[fill=red,draw=none] (1.452000,1.071120) circle (0.05);
\draw[-,black] (1.584000,0) -- (1.584000,1.332720);
\draw[fill=black,draw=none] (1.584000,1.332720) circle (0.05);
\draw[-,black] (1.716000,0) -- (1.716000,1.273320);
\draw[fill=black,draw=none] (1.716000,1.273320) circle (0.05);
\draw[-,black] (1.848000,0) -- (1.848000,1.219560);
\draw[fill=black,draw=none] (1.848000,1.219560) circle (0.05);
\draw[->] (-1.848000,0) -- (2.112000,0) node[at end, below] {};
\draw[thin,red] (-0.396000,1.114920) -- (-0.264000,1.160645);
\draw[thin,red] (-0.264000,1.160645) -- (-0.132000,1.198854);
\draw[thin,red] (-0.132000,1.198854) -- (0.000000,1.229547);
\draw[thin,red] (0.000000,1.229547) -- (0.132000,1.252724);
\draw[thin,red] (0.132000,1.252724) -- (0.264000,1.268385);
\draw[thin,red] (0.264000,1.268385) -- (0.396000,1.276531);
\draw[thin,red] (0.396000,1.276531) -- (0.528000,1.277160);
\draw[thin,red] (0.528000,1.277160) -- (0.660000,1.270273);
\draw[thin,red] (0.660000,1.270273) -- (0.792000,1.255871);
\draw[thin,red] (0.792000,1.255871) -- (0.924000,1.233953);
\draw[thin,red] (0.924000,1.233953) -- (1.056000,1.204518);
\draw[thin,red] (1.056000,1.204518) -- (1.188000,1.167568);
\draw[thin,red] (1.188000,1.167568) -- (1.320000,1.123102);
\draw[thin,red] (1.320000,1.123102) -- (1.452000,1.071120);
\draw[thin,red] (1.452000,1.071120) -- (1.584000,1.011622);
\draw[thin,red] (1.584000,1.011622) -- (1.716000,0.944608);
\draw[thin,red] (1.716000,0.944608) -- (1.848000,0.870078); 
  \end{tikzpicture}   
 }
\caption{Channel interpolation techniques using the estimates from (b) the actual slot, (c) the actual subframe and (d)-(e) the actual and previous subframe.}
\label{fig:mlerch_pilotssub} 
\end{figure}
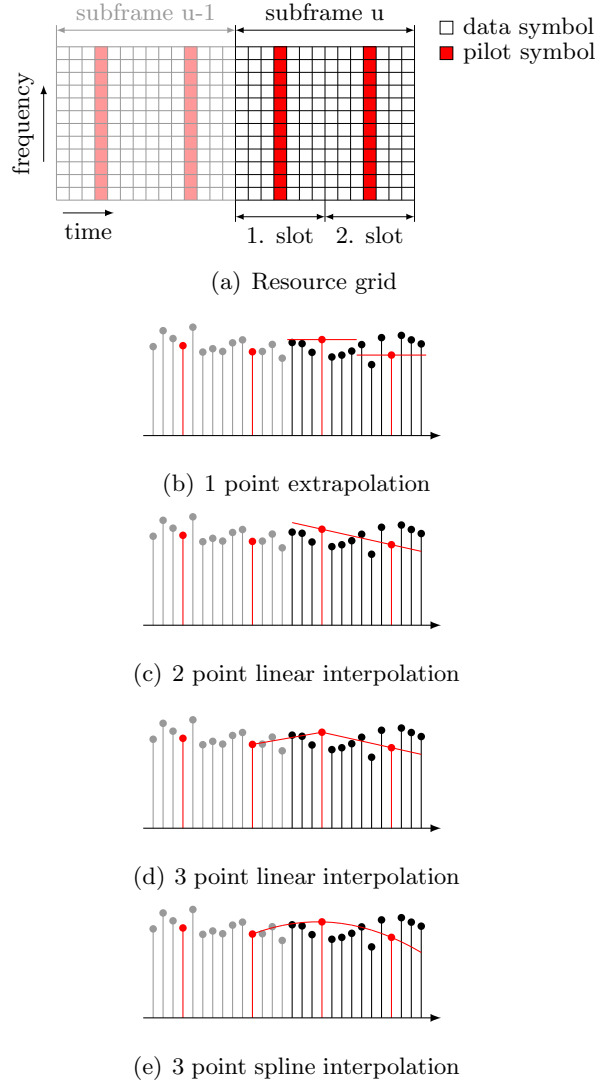

\begin{figure*}
\centering
\subfigure[Channel estimation and interpolation error at 30dB SNR.]{\includegraphics[width=0.45\textwidth]{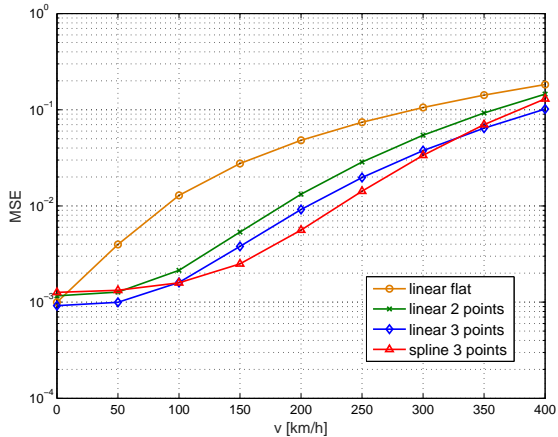}}
\subfigure[Uncoded {BER} for 4-QAM at 200 km/h.]{\includegraphics[width=0.45\textwidth]{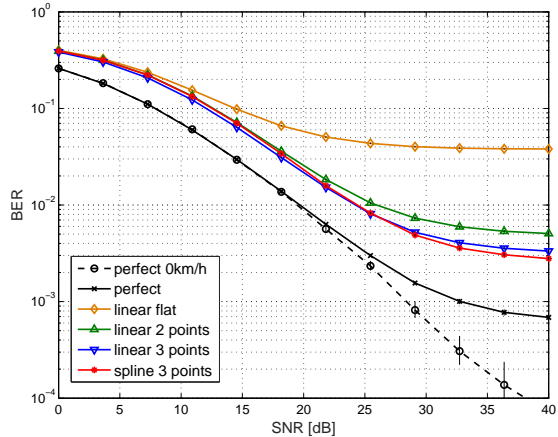}}
\caption{Comparison of channel interpolation techniques using different numbers of reference symbols and LS-SAV channel estimation.}
\label{fig:mlerch_Intresults} 
\end{figure*}

Under fast fading conditions additional effects influence the performance of {LTE} uplink transmissions. Doppler shifts degrade the {SINR} by introducing velocity dependent {ICI}~\cite{robertson1999effects} whereas the {SINR} increases with increasing subcarrier spacing. The subcarrier spacing of 15 kHz that is used in {LTE} makes transmissions quite robust against {ICI}. The impact of {ICI} becomes only evident at high velocities and high {SNR}. Fig.~\ref{fig:mlerch_Intresults}~(b) shows the {BER} for the case of perfect channel knowledge where the performance is only degraded by noise and {ICI}. At 200 km/h the {BER} saturates due to {ICI} at high {SNR} whereas {ICI} mitigation techniques~\cite{Nis15} show promising results to reduce this impact of {ICI}.\\

Another effect that hampers {LTE} transmissions at high velocities are temporal channel interpolation errors. While in the {LTE} downlink the pattern used to multiplex data and reference symbols is a good trad-off between a small temporal and spectral spacing accounting for highly frequency selective channels and fast fading channels and a rather small overhead, this is different in the uplink. As shown in Fig.~\ref{fig:mlerch_pilotssub}~(a) uplink {DMRS}s occupy the whole subband. While there is no need for interpolation over frequency, the temporal spacing is about twice the spacing of the reference symbols in the downlink. Furthermore, if frequency hopping is performed the number of adjacent pilots transmitted in the same subband is two for inter-subframe frequency hopping and only one for intra-subframe frequency hopping where frequency hopping is performed on a per-slot basis. Due to this special structure channel interpolation in the LTE uplink is a challenging problem. Therefore we investigated various channel interpolation techniques using a single, two or three consecutive pilot symbols. Fig.~\ref{fig:mlerch_pilotssub}~(b)-(e) illustrates the channel interpolation techniques considered. The highest channel interpolation errors (Fig.~\ref{fig:mlerch_Intresults}~(a)) are observed for \textit{1 point extrapolation} where the channel estimate obtained in a certain slot is used to equalize the symbols within that slot and no interpolation is performed at all. The higher the number of pilots involved in channel interpolation the lower the {MSE} gets. The results in terms of {BER} in Fig.~\ref{fig:mlerch_Intresults}~(b) show a similar behaviour.

For a measurement based comparison of interpolation techniques using channel estimates form both, the previous and the subsequent subframe the reader is referred to~\cite{Ler15}.


\section{Future Research Questions}
\label{sec:Future}

Until now our research efforts on the Vienna {LTE-A} Uplink Simulator have been concentrated on single links between user and base station, focusing on basic transceiver issues such as link adaptation and channel estimation. In future, our scope will shift to multi-user multi-base station scenarios, enabling on one hand exploitation of multi-user diversity in space, time and frequency and, on the other hand, consideration of interference in-between simultaneous transmissions from multiple base stations. 

We will address cross-layer multi-user scheduling, jointly optimizing multi-user resource allocation and per-user link adaptation; this is an intricate issue in {LTE}, due to the non-linear relationship between the resources assigned to a user and its corresponding {SC-FDM} {SINR}~(\ref{eq:SNR_SCFDMA}); we have already addressed this issue for the downlink in~\cite{Schwarz-Asilomar2010}. Multi-user scheduling, furthermore, has to find a favourable trade-off between transmission efficiency and fairness of resource allocation. We will extend \tempOff{our} existing downlink schedulers, which enable Pareto-efficient transmission with arbitrary fairness\tempOff{~\cite{schwarz2011throughput}}, to the uplink specifics and compare to other proposals, e.g.,~\cite{Survey_scheduling}.           

The realization of massive {MIMO} in {LTE} compliant systems is another highly important research topic, since it promises an order of magnitude network efficiency gains through spatial multiplexing of users~\cite{Rusek2013,Lu2014,Larsson2014}. Yet, many issues still need to be better understood and resolved to enable efficient massive {MIMO} transmission in practice. One important step towards reasonable performance investigation of massive antenna arrays is to employ realistic channel models, such as, the 3GPP three-dimensional channel model~\cite{tr36873}, which we plan to incorporate in future releases of our simulator.   

We finally plan to implement multi-base station support in future releases of the Vienna {LTE-A} Uplink Simulator. Even though, for reasons of computational complexity, simulations will be confined to comparatively small scenarios containing some few base stations, we still expect to extract valuable performance indicators for coordinated multipoint reception schemes~\cite{Coh_reception}, accounting for practical constraints, such as, limited back-haul capacity.           


\section{Conclusion}
\label{sec:Conclusion}

For an {LTE-A} uplink transmission model we derived {SINR} expressions, both with and without {DFT} pre-spreading. We specialized these equations to {ZF} and {MMSE} receivers and showed that ZF performance is strongly affected by the worst subcarrier. Comparing the resulting {BER} we  revealed, that {SC-FDM} performance is generally inferior to {OFDM} and that applying {MMSE} equalization is crucial to get closer to OFDM performance.

Based on the system's {SINR} we analysed the achievable rate. We also introduced a method to estimate the {SC-FDM} rate for $N_R > L$. Further a possible calculation of {LTE-A} link adaptation parameters was proposed to achieve throughout close to performance bounds.

Lastly, we considered methods to gather {CSI} at the receiver. We compared the performance of various channel estimation and interpolation  techniques. By incorporating the channel estimates of the previous subframe, we showed superior performance in terms of channel interpolation.


\section*{Appendix}

\subsection*{General MIMO SC-FDMA SINR expression}
\label{sec:AppendixA}

The signal estimates are described via the input-output relationship Equation (\ref{eq:IO}). We first slice out that part of $\Mat{K}$ which acts on layer $l$ by multiplying with the selector matrix $\Mat{S}^{(l)}$ from left. As indicated in (\ref{eq:IO}), the signal estimate consists of three contributions. 
\begin{itemize}[leftmargin=2.5cm] 
\item[signal:   ] $\hat{\Vec{x}}_{\rm s}= \Mat{S}^{(l)} \left(\Mat{I}\odot \Mat{K}\right) \Vec{x}$

\item[interference:   ] $\hat{\Vec{x}}_{\rm i}= \Mat{S}^{(l)}\left(\Mat{K}-\Mat{I}\odot \Mat{K}\right) \Vec{x}$

\item[noise:   ] $\hat{\Vec{x}}_{\rm n}=\Mat{S}^{(l)} \tilde{\Vec{n}} $
\end{itemize}

As $\Vec{x}$ and $\tilde{\Vec{n}}$ are zero mean random quantities, their power is described by means of the second moment. To calculate the second moments we take out the diagonal elements of the respective covariance matrices of each contribution.

\begin{eqnarray}
\hspace{-12ex}&{{\rm SINR}_l^{\rm SC-FDM}} = \\ \hspace{-12ex}& \bigg[\left(\!\Mat{I}\odot\mathbb{E}\lbrace\hat{\Vec{x}}_{\rm s} \hat{\Vec{x}}_{\rm s}^H \rbrace \!\right)\left(\!{\Mat{I}\odot \mathbb{E}\lbrace \hat{\Vec{x}}_{\rm i} \hat{\Vec{x}}_{\rm i}^H  \rbrace+\Mat{I}\odot\mathbb{E}\lbrace\hat{\Vec{x}}_{\rm n} \hat{\Vec{x}}_{\rm n}^H  \rbrace} \! \right)^{-1} \bigg]_{1,1} \notag
\end{eqnarray}

Before we derive the different covariance matrices, we recapitulate a required property of circulant matrices.
A circulant matrix $\Mat{C} \in \mathbb{C}^{N \times N}$ is fully described by its first column $\Vec{c}$, as its eigenvectors are the DFT basis-vectors and its eigenvalues are the {DFT} of $\Vec{c} = (c_0, c_{1}, \dots, c_{N-1}) $.
\begin{eqnarray}
\Mat{C} &=\left(\begin{array}{ccccc}
c_0     & c_{N-1} &  \dots & c_{1}  \\
c_{1} & c_0    &          & c_{2}  \\
\vdots  & &  \ddots  & \vdots   \\
c_{N-1}  & \dots &  c_{1} & c_0 \\
\end{array}\right) \\
{} \nonumber\\
&=\Mat{D}^H \operatorname{Diag} \left(\Mat{D}\Vec{c} \right)\Mat{D} = \Mat{D}^H \Mat{\Lambda} \Mat{D}
\end{eqnarray}
The main diagonal elements $c_0$ of $\Mat{C}$ are given by 
\be \label{eq:diag_elem}
c_0=\frac{1}{N}\sum_{i=0}^{N-1} [\Mat{D}\Vec{c}]_{i} = \frac{1}{N}\sum_{i=0}^{N-1} [\Mat{\Lambda}]_{i,i} =\frac{1}{N}\mathds{1}^T \diag \!\!\left(\Mat{\Lambda}\right)~.
\ee  \\

\subsubsection*{$\mathbb{E}\lbrace\hat{\Vec{x}}_{\rm s} \hat{\Vec{x}}_{\rm s}^H \rbrace$:}

The input-output matrix $\Mat{K}$ is of block-circulant structure, as illustrated in Figure \ref{fig:channels}~(c). The eigenvalues of the diagonal blocks are given by  $\diag \!\! \left(\Mat{\Lambda}\right)=\Mat{S}^{(l)} \diag \!\!\!\left(\Mat{F} \Mat{H}_{\rm eff} \right)$ and the diagonal elements of the $l^{\rm th}$ diagonal block are then $\frac{1}{N}\mathds{1}^T \Mat{S}^{(l)} \diag \!\!\!\left(\Mat{F}\Mat{H}_{\rm eff}  \right)$ as asserted by Equation (\ref{eq:diag_elem}), thus 
\be 
\Mat{S}^{(l)} \left(\Mat{I}\odot \Mat{K} \right) = \frac{1}{N}\mathds{1}^T \Mat{S}^{(l)} \diag \!\!\! \left( \Mat{F}\Mat{H}_{\rm eff} \right) \Mat{I}
\ee
Assuming zero mean,white data with variance $\sigma_x^2$ the diagonal elements of $\mathbb{E}\lbrace\hat{\Vec{x}}_{\rm s} \hat{\Vec{x}}_{\rm s}^H \rbrace$ are given by $\sigma_x^2| \frac{1}{N} \mathds{1}^T \Mat{S}^{(l)} \diag \!\!\! \left( \Mat{F}\Mat{H}_{\rm eff}\right)|^2$.

\subsubsection*{$\mathbb{E}\lbrace\hat{\Vec{x}}_{\rm i} \hat{\Vec{x}}_{\rm i}^H \rbrace$:}

If $\Mat{C}$ is circulant 
\be
\tilde{\Mat{C}}=\Mat{C}-c_0 \Mat{I} = \left(\begin{array}{ccccc}
0     & c_{N-1} &  \dots & c_{1}  \\
c_{1} & 0    &          & c_{2}  \\
\vdots  & &  \ddots  & \vdots   \\
{c_{N-1} } & \dots &  c_{1} & 0 \\
\end{array} \right)
\ee
is circulant as well and the diagonal elements of $\tilde{\Mat{C}} \tilde{\Mat{C}}^H$ are the sum of the magnitude squares of $\tilde{\Vec{c}} = (0, c_{1}, \dots, c_{N-1}) $. Using Parseval's theorem we arrive at
\begin{eqnarray} \label{eq:intra_inf} 
\sum_{i=1}^{N-1} |c_i|^2 =& \frac{1}{N} \sum_{j=1}^{N-1} |[\Mat{\Lambda}]_{j,j}|^2 \\ =& \frac{1}{N}\sum_{j=0}^{N-1} |[\Mat{\Lambda}]_{j,j}|^2 -  \bigg| \frac{1}{N}\sum_{j=0}^{N-1} [\Mat{\Lambda}]_{j,j}\bigg|^2~. \notag
\end{eqnarray}

\noindent The inter-layer interference consists of $L-1$ $\Mat{C}$-type blocks, where we simply average the magnitude squares of the eigenvalues, i.e., the corresponding block-part of $\Mat{F}\Mat{H}_{\rm eff}$. The intra-layer interference is described via a $\tilde{\Mat{C}}$ block and is given in  Equation (\ref{eq:intra_inf}). Both contributions can be compactly written as
\be 
\sigma_x^2 \frac{1}{N}\| \Mat{S}^{(l)} \Mat{F}\Mat{H}_{\rm eff} \|_F^2 - \sigma_x^2 \big| \frac{1}{N} \mathds{1}^T \Mat{S}^{(l)} \diag \!\!\! \left(  \Mat{F}\Mat{H}_{\rm eff} \right) \! \big|^2~. 
\ee

\subsubsection*{$\mathbb{E}\lbrace\hat{\Vec{x}}_{\rm n} \hat{\Vec{x}}_{\rm n}^H  \rbrace$:}
The noise covariance matrix is circulant as well and the detailed derivations can be found in \cite{zoechmann:mimotransmission}.

\subsection*{SISO MMSE SC-FDMA SINR expression}

For a SISO system and an one-tap equalizer the expression $\Mat{F}\Mat{H}_{\rm eff}$ is of diagonal shape. \cite{zoechmann:mimotransmission} has shown, that the MMSE equalizer for SC-FDM equals the OFDM expression, i.e., $\Mat{F}=(  \frac{\sigma_n^2}{\sigma_x^2} \Mat{I} + \Mat{H}_{\rm eff}^H \Mat{H}_{\rm eff})^{-1} \Mat{H}_{\rm eff}^H$. Thus, the elements on the main diagonal of $\Mat{F}\Mat{H}_{\rm eff}$ are simply given by $|\Mat{H}_k|^2(\frac{\sigma_n^2}{\sigma_x^2}+|\Mat{H}_k|^2)^{-1}$ and we rewrite (\ref{eq:SNR_SCFDMA}) to (\ref{eq:SISO_der}), where we have used the identity

\begin{scriptsize}
\be 
\frac{1}{N_{SC}}  \sum\limits_{k=1}^{N_{SC}} \frac{|\Mat{H}_k|^2}{\frac{\sigma_n^2}{\sigma_x^2}+|\Mat{H}_k|^2} =1 - \frac{\sigma_n^2}{\sigma_x^2} \frac{1}{N_{SC}}  \sum\limits_{k=1}^{N_{SC}} \frac{1}{\frac{\sigma_n^2}{\sigma_x^2}+|\Mat{H}_k|^2}
\ee
\end{scriptsize}

\noindent from \cite{sanchez2011ber}.

\begin{table*}
\begin{eqnarray}
{\rm SINR}_{\rm MMSE}^{\rm SC-FDM} &= \frac{\frac{\sigma_x^2}{N_{SC}}  \left(\sum\limits_{k=1}^{N_{SC}} \frac{|\Mat{H}_k|^2}{\frac{\sigma_n^2}{\sigma_x^2}+|\Mat{H}_k|^2}\right)^2}{\sigma_x^2 \sum\limits_{k=1}^{N_{SC}} \left(\frac{|\Mat{H}_k|^2}{\frac{\sigma_n^2}{\sigma_x^2}+|\Mat{H}_k|^2}\right)^2-\frac{\sigma_x^2}{N_{SC}}  \left(\sum\limits_{k=1}^{N_{SC}} \frac{|\Mat{H}_k|^2}{\frac{\sigma_n^2}{\sigma_x^2}+|\Mat{H}_k|^2}\right)^2 +\sigma_n^2 \sum\limits_{k=1}^{N_{SC}} \frac{|\Mat{H}_k|^2}{\left(\frac{\sigma_n^2}{\sigma_x^2}+|\Mat{H}_k|^2\right)^2 }}  \\
&= \frac{\frac{1}{N_{SC}}  \left(\sum\limits_{k=1}^{N_{SC}} \frac{|\Mat{H}_k|^2}{\frac{\sigma_n^2}{\sigma_x^2}+|\Mat{H}_k|^2}\right)^2}{\left( \sum\limits_{k=1}^{N_{SC}} \frac{|\Mat{H}_k|^2}{\frac{\sigma_n^2}{\sigma_x^2}+|\Mat{H}_k|^2 }\right) -\frac{1}{N_{SC}}  \left(\sum\limits_{k=1}^{N_{SC}} \frac{|\Mat{H}_k|^2}{\frac{\sigma_n^2}{\sigma_x^2}+|\Mat{H}_k|^2}\right)^2} = \frac{\frac{1}{N_{SC}}  \sum\limits_{k=1}^{N_{SC}} \frac{|\Mat{H}_k|^2}{\frac{\sigma_n^2}{\sigma_x^2}+|\Mat{H}_k|^2}}{ 1 -\frac{1}{N_{SC}} \sum\limits_{k=1}^{N_{SC}} \frac{|\Mat{H}_k|^2}{\frac{\sigma_n^2}{\sigma_x^2}+|\Mat{H}_k|^2}} \\ &= \frac{\sigma_x^2}{\sigma_n^2}\frac{1- \frac{\sigma_n^2}{\sigma_x^2} \frac{1}{{N_{SC}}}   \sum\limits_{k=1}^{N_{SC}} \frac{1}{\frac{\sigma_n^2}{\sigma_x^2}+|\Mat{H}_k|^2}    }{\frac{1}{{N_{SC}}}\sum\limits_{k=1}^{N_{SC}} \frac{1}{\frac{\sigma_n^2}{\sigma_x^2}+|\Mat{H}_k|^2}} \label{eq:SISO_der}
\end{eqnarray}
\hrule
\end{table*}


\bibliography{mybib}   
\bibliographystyle{IEEEtran}

\end{document}